\documentstyle[11pt,aaspp4,flushrt]{article}

\lefthead{Siegler \& Riffert}
\righthead{Relativistic SPH with artificial viscosity}

\newcommand{\etal}{et al.~}
\newcommand{\figuresize}{0.41\textwidth}
\newcommand{\boxsize}{0.89\textwidth}

\begin{document}

\title{Smoothed Particle Hydrodynamics Simulations of\\
Ultra-Relativistic Shocks with Artificial Viscosity} 

\author{S. Siegler and H. Riffert}

\affil{Institut f\"ur Astronomie und Astrophysik, Universit\"at
T\"ubingen \\ Auf der Morgenstelle 10, D-72076 T\"ubingen, Germany \\ 
Email: siegler,riffert@tat.physik.uni-tuebingen.de}

\authoremail{siegler,riffert@tat.physik.uni-tuebingen.de}

\begin{abstract}
We present a fully Lagrangian conservation form of the general relativistic
hydrodynamic equations for perfect fluids with artificial viscosity in a
given arbitrary background spacetime. This conservation formulation is
achieved by choosing suitable Lagrangian time evolution variables, from which
the generic fluid variables rest-mass density, 3-velocity, and thermodynamic
pressure have to be determined. We present the corresponding equations for an
ideal gas and show the existence and uniqueness of the solution. On the basis
of the Lagrangian formulation we have developed a three-dimensional general
relativistic Smoothed Particle Hydrodynamics (SPH) code using the standard
SPH formalism as known from non-relativistic fluid dynamics. One-dimensional
simulations of a shock tube and a wall shock are presented. With our method
we can model ultra-relativistic fluid flows including shocks with
relativistic $\gamma$-factors of even $1000$.
\end{abstract}
\section{INTRODUCTION}
Modeling ultra-relativistic fluid flows is a great challenge for any
relativistic hydro code. Numerical difficulties arise from strong
relativistic shocks and from narrow physical structures (Norman \& Winkler
1986). Typical examples in astrophysics for such extreme conditions are
proto-stellar jets and blast waves of supernovae explosions. In recent years,
the development of numerical algorithms for relativistic fluid dynamics went
mainly along two different lines. First, there are the so called High
Resolution Shock Capturing (HRSC) methods, which allow to obtain numerically
discontinuous solutions of the relativistic hydrodynamic equations by solving
local Riemann problems between adjacent numerical cells. Some recently
developed HRSC techniques are those of Font \etal (1994), Falle \& Komissarov
(1996), Romero \etal (1996), Banyuls \etal (1997), Wen, Panaitescu, \& Laguna
(1997), Pons \etal (1998), and Komissarov (1999). However, employing analytic
solutions of the Riemann shock tube problem, it is not surprising that these
HRSC codes produce almost exact numerical solutions of flow structures
including discontinuities. In the second type of algorithms, shocks are
treated numerically not as fluid discontinuities, but are rather spread over
some small length with the help of an artificial viscosity. These algorithms
either solve the dynamical equations of relativistic hydrodynamics on an
Eulerian grid such as the finite difference schemes of Hawley, Smarr, \&
Wilson (1984b) and Norman \& Winkler (1986) or by using computational nodes
following the fluid motion such as the Lagrangian Smoothed Particle
Hydrodynamics (SPH) methods of Kheyfets, Miller, \& Zurek (1990) and Laguna,
Miller, \& Zurek (1993). All the latter methods seem to be
limited even for mildly relativistic flows containing shocks. Norman \&
Winkler (1986) suggest that the appearance of numerical inaccuracies and
instabilities is due to the time derivative of the relativistic
$\gamma$-factor in the energy equation of these numerical schemes. This
additional time derivative of a hydrodynamic variable renders the system of
evolution equations non-conservative. For non-relativistic hydrodynamics, it
is well-known that a numerical method based on non-conservative equations can
produce a solution which looks reasonable but is entirely wrong if shocks or
other discontinuities are involved (see LeVeque 1997 for a corresponding
analysis of Burger's equation).

The main intention of this paper is to present a conservation formulation of
the relativistic hydrodynamic equations that is designed for numerical
methods. This particular formulation allows hydro codes to resolve
ultra-relativistic shocks numerically with relativistic $\gamma$-factors of
even $1000$ by means of an artificial viscosity rather than using a Riemann
solver. The conservation form of the relativistic equations is obtained by
choosing suitable Lagrangian variables. Unfortunately, these Lagrangian
dynamical variables can be expressed in terms of the generic fluid variables
rest-mass density, 3-velocity, and thermodynamic pressure only through a set
of nonlinear algebraic equations. However, we show that these equations can
be solved analytically in a unique way.

For numerical work we discretize our set of partial differential equations by
the Smoothed Particle Hydrodynamics (SPH) method introduced by Lucy (1977)
and Gingold \& Monaghan (1977). In recent years, SPH became a popular
computational tool for numerically modeling complex three-dimensional fluid
flows in astrophysics. This is primarily due to its computational simplicity
and the absence of a computational grid. Furthermore, SPH is adaptive in the
sense that its computational nodes follow the fluid.

SPH has been already applied to relativistic fluid flows. Kheyfets \etal
(1990) developed a relativistically covariant version of the SPH technique
by modeling the contact interactions with spatial smoothing functions in the
local comoving frame of the fluid. Laguna \etal (1993) applied the SPH method
to the so called ADM formalism of general relativity due to Arnowitt, Deser,
\& Misner (1962). In their relativistic SPH formulation they modified the
flat space kernels of the Newtonian SPH method which can become anisotropic
and are then no longer invariant under translations which leads to additional
terms in the SPH equations. In contrast to these methods, we have developed a
fully three-dimensional general relativistic SPH code on the basis of our set
of Lagrangian conservation equations. This code employs the standard SPH
approach as used in Newtonian theory with spherically symmetric kernels for
all particles. Our code is restricted to ideal fluids with artificial
viscosity using the ideal-gas equation of state. The influence of the fluid
on the spacetime metric is neglected, therefore, we consider only a
background spacetime with a given but otherwise arbitrary metric. In
addition, we neglect the fluid's self-gravity and do not account for
radiation or electromagnetic effects.

The layout of this paper is as follows. In section \ref{hydroeq} we derive
the formulation of the relativistic hydrodynamic equations in Lagrangian
conservative form and present the equations for calculating the generic fluid
properties rest-mass density, 3-velocity, and thermodynamic pressure from our
suitably chosen Lagrangian variables. Section \ref{spheq} gives a review of
the standard SPH method that we use, including a prescription of its
application to relativistic fluid flows in curved spacetimes and the
implementation of an artificial viscosity. Numerical test calculations of
one-dimensional ultra-relativistic flows with discontinuities are presented
in section \ref{codetests}. Throughout this paper we set the speed of light
$c$ and the Boltzmann's constant $k$ to unity, i.e., $c=k=1$. Latin indices
$\{i,j,\dots\}$ run from $1$ to $3$, Greek indices $\{\mu,\nu,\dots\}$ from
$0$ to $3$; the signature of the metric is $(-,+,+,+)$.
\section{\label{hydroeq}THE LAGRANGIAN CONSERVATION FORMULATION OF THE
GENERAL RELATIVISTIC IDEAL FLUID EQUATIONS}
Our starting point is the covariant formulation of the equations of
relativistic hydrodynamics for a perfect fluid with artificial viscosity: the
local conservation of baryon number
\begin{equation}
\label{covconti}
\left(\rho u^\mu\right)_{;\mu} = 0
\end{equation}
and the local conservation of energy-momentum
\begin{equation}
\label{covenmom}
T^{\mu\nu}_{\enspace\enspace;\nu} = 0 \enspace
\end{equation}
with the stress-energy tensor of a perfect fluid
\begin{equation}
\label{stressenergy}
T^{\mu\nu}=(\rho w+q)u^\mu u^\nu +(p+q)g^{\mu\nu} \enspace .
\end{equation}
The semicolon denotes the covariant derivative and the Einstein summation
convention is used. Here, $\rho$ is the rest-mass density, $u^\mu$ the
4-velocity of the fluid, $p$ the isotropic thermodynamic pressure,
$w=1+\varepsilon+p/\rho$ the relativistic specific enthalpy with specific
internal energy $\varepsilon$, $q$ the artificial viscous pressure, and
$g^{\mu\nu}$ the spacetime metric. All thermodynamic quantities in the
stress-energy tensor (\ref{stressenergy}) are measured in the local rest
frame of the fluid. The artificial viscous pressure $q$, which appears in
equation (\ref{stressenergy}), can be introduced into the stress-energy
tensor either by the substitution $p\rightarrow p+q$ or, equivalently, from
the stress-energy tensor of a viscous fluid (Misner, Thorne, \& Wheeler 1973;
Landau \& Lifshitz 1991) by ignoring the shear viscosity and heat conduction
and replacing the bulk viscous pressure by $q$. An explicit expression for
$q$ in terms of velocity gradients will be given in the subsequent section.

For a Lagrangian formulation of relativistic hydrodynamics suitable for SPH
one has to break the unity of time and space inherent in the covariant
formulation. This can be accomplished by applying the ADM formalism of
Arnowitt \etal (1962), where the spacetime is decomposed into an infinite
foliation of spatial hypersurfaces $\Sigma_t$ of constant coordinate time $t$
by writing the line element as
\begin{displaymath}
ds^2 = g_{\mu\nu} dx^\mu dx^\nu = -\left(\alpha^2 - \beta^i\beta_i\right)dt^2
 + 2\beta_i dx^i + \eta_{ij} dx^i dx^j \enspace .
\end{displaymath}
Here, $\alpha$ is the {\it lapse} function, $\beta^i$ the {\it shift} vector,
and $\eta_{ij}$ the spatial metric induced on $\Sigma_t$ with $\beta_i =
\eta_{ij}\beta^j$ and $\eta_{il}\eta^{lj} = \delta_{i}^{~j}$. In this paper
we consider only given background spacetimes, i.e., we do not solve the
Einstein equations. Thus, $g_{\mu\nu}$, $\alpha$, $\beta^i$, and $\eta_{ij}$
are given analytic functions of both space and time. From the definitions of
$\alpha$ and $\beta^i$ the basis vector field $\mbox{\boldmath$\partial$}_t$
of the coordinate basis
$\{\mbox{\boldmath$\partial$}_t,\mbox{\boldmath$\partial$}_i\}$ can be
decomposed into normal and parallel components with respect to the
hypersurfaces $\Sigma_t$
\begin{equation}
\label{partialt}
\mbox{\boldmath$\partial$}_t = \alpha \mbox{\boldmath$n$} + \beta^i
\mbox{\boldmath$\partial$}_i \enspace ,
\end{equation}
where $\mbox{\boldmath$n$}$ is the unit time-like vector field normal to the
slices $\Sigma_t$, i.e., $\mbox{\boldmath$n\cdot\partial$}_i=0$. Observers
having $\mbox{\boldmath$n$}$ as 4-velocity are at rest in the slices
$\Sigma_t$ --- they are called {\it Eulerian observers}. In the basis
$\{\mbox{\boldmath$n$},\mbox{\boldmath$\partial$}_i\}$ the 4-velocity of a
fluid has the representation
\begin{displaymath}
\mbox{\boldmath$u$} = \gamma \left( \mbox{\boldmath$n$} + \bar v^i
\mbox{\boldmath$\partial$}_i \right) \enspace ,
\end{displaymath}
whereas in the coordinate basis
$\{\mbox{\boldmath$\partial$}_t,\mbox{\boldmath$\partial$}_i\}$ it follows from
equation (\ref{partialt})
\begin{displaymath}
\mbox{\boldmath$u$} = u^\mu\mbox{\boldmath$\partial_\mu$} =
\frac{\gamma}{\alpha} \left( \mbox{\boldmath$\partial$}_t + v^i
\mbox{\boldmath$\partial$}_i \right)
\end{displaymath}
with
\begin{displaymath}
v^i = \alpha \bar v^i - \beta^i \enspace .
\end{displaymath}
From the normalization condition of the 4-velocity $u^\mu u_\mu=-1$
the relativistic $\gamma$-factor is given by
\begin{displaymath}
\gamma = \frac{1}{\sqrt{1-\eta_{ij}{\bar v^i}{\bar v^j}}} \enspace .
\end{displaymath}
In the following, we use both, the 3-velocity of the fluid ${\bar v^i}$
measured by Eulerian observers and the 3-velocity $v^i$ in the coordinate
basis.

We now derive the Lagrangian equations of relativistic hydrodynamics where
the Lagrangian or total time derivative $d/dt$ is defined as 
\begin{displaymath}
\frac{d}{dt} = \frac{\alpha}{\gamma} u^\mu \partial_\mu =
\partial_t+v^i\partial_i \enspace .
\end{displaymath}
From the law of baryon-number conservation~(\ref{covconti}) we obtain
\begin{eqnarray}
\label{lagconti}
0 & = & \partial_\mu\left(\sqrt{-g}\rho
u^\mu\right)=\partial_t\left(\sqrt{-g}\rho\frac{\gamma}{\alpha}\right) +
\partial_i\left(\sqrt{-g}\rho \frac{\gamma}{\alpha}v^i\right)\nonumber\\
& = & \frac{d D^*}{dt} + D^*\partial_iv^i \enspace ,
\end{eqnarray}
where $g$ is the determinant of the spacetime metric $g^{\mu\nu}$, and
the relativistic rest-mass density $D^*$ is defined by
\begin{equation}
\label{densityvar}
D^* = \sqrt{-g}\frac{\gamma}{\alpha}\rho = \sqrt{\eta}\gamma\rho
\enspace .
\end{equation}
Here, $\eta$ is the determinant of the spatial metric $\eta_{ij}$ obeying
$\sqrt{\eta} = \sqrt{-g}/\alpha$. With the above definition of $D^*$ the
relativistic continuity equation (\ref{lagconti}) has the same form as in the
non-relativistic case. Note, however, that in equation (\ref{lagconti}) the
expression for $\partial_i v^i$ is not the divergence of a 3-vector on
$\Sigma_t$ but rather a sum of partial derivatives. Eulerian observers
measure the relativistic rest-mass density $D = -\rho\mbox{\boldmath$u\cdot
n$} = \rho\gamma$. By rewriting equation (\ref{lagconti}) for $D$, we obtain
the relativistic continuity equation
\begin{displaymath}
0 = \frac{dD}{dt} + D\partial_iv^i + D\frac{d}{dt}\ln{\sqrt{\eta}} 
\end{displaymath}
which, in contrast to equation (\ref{lagconti}), contains an additional
source term. In this paper we will use both definitions of a relativistic
rest-mass density, i.e., $D$ and $D^*=\sqrt{\eta}D$.

In order to derive our relativistic expressions for the energy and momentum
equations, we first re-express the conservation law of energy-momentum
(\ref{covenmom}) by using the continuity equation (\ref{covconti})
\begin{equation}
\label{covenmom2}
0={T_\mu^{\enspace\nu}}_{;\nu} = \rho
u^\nu\left[\left(w+\frac{q}{\rho}\right)u_\mu\right]_{;\nu} +
\partial_\mu(p+q)\enspace .
\end{equation}
One can easily show that the covariant derivative in the first term of
equation (\ref{covenmom2}) can be written as
\begin{displaymath}
u^\nu \left[\left(w+\frac{q}{\rho}\right)u_\mu\right]_{;\nu}
=\frac{\gamma}{\alpha}\frac{d}{dt}\left[\left(w+\frac{q}{\rho}\right)u_\mu
\right] - \frac{1}{2}\left(w+\frac{q}{\rho}\right)g_{\alpha\beta,\mu}
u^\alpha u^\beta \enspace .
\end{displaymath}
Thus, equation (\ref{covenmom2}) becomes
\begin{eqnarray}
\label{covenmom3}
\frac{d}{dt}\left[\left(w+\frac{q}{\rho}\right)u_\mu\right]& = &
-\frac{\alpha}{\rho\gamma}\left[\partial_\mu(p + q) - \frac{1}{2}(\rho
w+q)u^\alpha u^\beta g_{\alpha\beta,\mu}\right]\nonumber\\ 
& = & -\frac{1}{D^*}\left[\partial_\mu\left[\sqrt{-g}(p+q)\right] -
\frac{\sqrt{-g}}{2}T^{\alpha\beta}g_{\alpha\beta,\mu}\right] \enspace ,
\end{eqnarray}
where we have used the identity
\begin{displaymath}
\frac{1}{\sqrt{-g}}\partial_\mu\left(\sqrt{-g}\right) = \frac{1}{2}
g^{\alpha\beta}g_{\alpha\beta,\mu} \enspace .
\end{displaymath}

Taking the spatial components of equation (\ref{covenmom3}), i.e.,
$\mu=i$, and using 
\begin{displaymath}
u_i = g_{i\mu}u^\mu = \beta_i\frac{\gamma}{\alpha} + \eta_{ij}\frac{\gamma}
{\alpha}v^j = \gamma\eta_{ij}\bar v^j \enspace ,
\end{displaymath}
we get the momentum equation
\begin{equation}
\label{lagmomentum}
\frac{d}{dt}S_i = -\frac{1}{D^*}\left[\partial_i\left[\sqrt{-g}(p+q)\right] -
\frac{\sqrt{-g}}{2}T^{\alpha\beta}\partial_i g_{\alpha\beta}\right] \enspace ,
\end{equation}
with the relativistic specific momentum $S_i$ defined by
\begin{equation}
\label{momentumvar}
S_i = \left(w+\frac{q}{\rho}\right)\gamma\eta_{ij}\bar v^j \enspace .
\end{equation}
The component $DS_i = - \mbox{\boldmath$T$} (\mbox{\boldmath$n$} ,
\mbox{\boldmath$\partial$}_i)$ of the stress-energy tensor field
$\mbox{\boldmath$T$}$ is the relativistic momentum density in the
$i$-direction measured by Eulerian observers. The momentum equation
(\ref{lagmomentum}), which, in a similar form, was also used by Laguna \etal
(1993), can be applied immediately to the SPH method because it contains only
spatial derivatives on its right hand side. In the Eulerian formulation an
expression similar to equation (\ref{lagmomentum}) without artificial
viscosity is given by Hawley, Smarr, \& Wilson (1984a) for the momentum
variable $DS_i$, which is more convenient for a Eulerian description. The
Newtonian limit yielding the non-relativistic Euler equation is obvious from
equation (\ref{lagmomentum}).

Taking the $\mu=0$--component of equation (\ref{covenmom3}), we obtain the
relativistic energy equation
\begin{equation}
\label{covenmom4}
\frac{d}{dt}\left[\left(w + \frac{q}{\rho}\right)u_0\right] =
-\frac{1}{D^*}\left[\partial_t\left[\sqrt{-g}(p + q)\right] -
\frac{\sqrt{-g}}{2} T^{\alpha\beta}\partial_t g_{\alpha\beta}\right]
\enspace ,
\end{equation}
which, unfortunately, has time derivatives of hydrodynamic variables on both
sides. Re-expressing the left hand side of equation (\ref{covenmom4}) as
\begin{displaymath}
\left(w + \frac{q}{\rho}\right)u_0 = -\alpha\left(w +
\frac{q}{\rho}\right)\gamma + \beta^iS_i
\end{displaymath}
with 
\begin{displaymath}
u_0 = g_{0\mu}u^\mu = \left(\beta^i\beta_i-\alpha^2\right)
\frac{\gamma}{\alpha} + \beta_i\frac{\gamma}{\alpha} v^i
= \gamma\left(\eta_{ij}\beta^i\bar v^j - \alpha\right)
\enspace ,
\end{displaymath}
and rewriting the first term on the right hand side of equation
(\ref{covenmom4}) using equation (\ref{lagconti})\,,
\begin{eqnarray*}
\frac{1}{D^*} \partial_t\left[\sqrt{-g}(p + q)\right] & = &
\frac{1}{D^*}\frac{d}{dt}\left[\sqrt{-g}(p + q)\right] -
\frac{v^i}{D^*}\partial_i\left[\sqrt{-g}(p + q)\right]\nonumber\\
& = & \frac{d}{dt}\left(\sqrt{-g}\frac{p+q}{D^*}\right)
- \frac{1}{D^*}\partial_i\left[\sqrt{-g}(p + q)v^i\right]
\enspace ,
\end{eqnarray*}
we obtain the relativistic energy equation
\begin{equation}
\label{lagenergy}
\frac{d}{dt}\left[\alpha E - \beta^iS_i\right] =
-\frac{1}{D^*}\left[\partial_i\left[\sqrt{-g}(p + q)v^i\right] +
\frac{\sqrt{-g}}{2}T^{\alpha\beta}\partial_tg_{\alpha\beta}\right] \enspace ,
\end{equation}
with the total relativistic specific energy $E$ defined as
\begin{equation}
\label{energyvar}
E = \left(w + \frac{q}{\rho}\right)\gamma - \frac{p + q}{D} \enspace .
\end{equation}
The component $DE = \mbox{\boldmath$T$} (\mbox{\boldmath$n$},
\mbox{\boldmath$n$})$ of the stress-energy tensor field $\mbox{\boldmath$T$}$
is the total relativistic energy density measured by Eulerian observers.
As in the case of the relativistic momentum equation (\ref{lagmomentum}) the
right hand side of equation (\ref{lagenergy}) contains no time derivatives of
hydrodynamic variables. It is, therefore, well suited for the SPH method in
contrast to the energy equation used by Hawley \etal (1984a, 1984b) and
Laguna \etal (1993) which has a non-conservative form containing two total
time derivatives of hydrodynamical variables separately on both sides of the
equation. Norman \& Winkler (1986) suggest that the additional time
derivative of the relativistic $\gamma$-factor in the energy equation gives
rise to numerical inaccuracies and instabilities. Note that for the special
relativistic case equation (\ref{lagenergy}) without an artificial viscosity
can also be found in Monaghan (1992). In the non-relativistic case, equation
(\ref{lagenergy}) yields the energy equation for the total non-relativistic
specific energy $|\mbox{\boldmath$v$}|^2/2+\varepsilon_N$ written in a form
similar to our expression (\ref{lagenergy})
\begin{displaymath}
\frac{d}{dt}\left[\frac{1}{2} |\mbox{\boldmath$v$}|^2 + \left(w_N +
\frac{q}{\rho_N}\right) - \frac{p + q}{\rho_N}\right] =
-\frac{1}{\rho_N}\partial_i\left[(p + q)v^i\right] \enspace ,
\end{displaymath}
where the index $N$ denotes Newtonian quantities and $w_N = \varepsilon_N +
p/\rho_N$ is the non-relativistic specific enthalpy.

To close our system of hydrodynamical equations (\ref{lagconti}),
(\ref{lagmomentum}), and (\ref{lagenergy}), we have to add an equation of
state of the form $p=p(\rho,\varepsilon)$, which relates the thermodynamic
pressure $p$ to the rest-mass density $\rho$ and the specific internal energy
$\varepsilon$. We restrict ourselves to the ideal-gas equation of state given
by
\begin{equation}
\label{idealstateeq}
p = (\Gamma - 1)\rho\varepsilon
\end{equation}
with the ideal-gas adiabatic constant $\Gamma$.

Our system of ideal fluid equations is now complete. We have derived the
relativistic hydrodynamic equations (\ref{lagconti}), (\ref{lagmomentum}),
and (\ref{lagenergy}) in Lagrangian conservative form similar to their
Newtonian counterparts. This was achieved by choosing suitable hydrodynamic
variables $D^*$, $S_i$, and $E$ defined in equations (\ref{densityvar}),
(\ref{momentumvar}), and (\ref{energyvar}). Because of the equation of state
(\ref{idealstateeq}) and the use of the 3-velocity for moving particles in
the Lagrangian numerical methods, we now need to calculate the generic
hydrodynamic quantities $\rho$, $\bar v^i$, and $p$ from these variables by
solving a highly nonlinear system of equations. If an artificial viscous
pressure $q$ is included, a severe difficulty arises from $q$ being usually
expressed in terms of velocity gradients. Since there is no time evolution
equation for $q$, the character of the dynamic equations is thus changed
which is the usual situation in the hydrodynamics of viscous flows. Due to
the coupling of $\bar v^i$ and $q$, the suitably chosen dynamic variables
actually have to be solved iteratively for the generic hydrodynamic
variables. However, artificial viscosity operates only in the vicinity of
shock transitions and is zero everywhere else. For simplicity one can
calculate the generic variables in a single iteration taking $q$ from the
previous time step. Using the expression $w=1+p/(G\rho)$ with $G=1-1/\Gamma$
for the total relativistic specific enthalpy, the specific energy $E$ from
equation (\ref{energyvar}) can be written as
\begin{eqnarray}
\label{energyvar2}
E & = & w\gamma - (w-1)\frac{G}{\gamma} + \left(\gamma -
\frac{1}{\gamma}\right)\frac{q}{\rho}\nonumber\\ 
& = & \left(\gamma - \frac{G}{\gamma}\right)w + \frac{G}{\gamma} +
\left(\gamma - \frac{1}{\gamma}\right)\frac{q}{\rho}\enspace .
\end{eqnarray}
Solving equation (\ref{energyvar2}) for the relativistic specific enthalpy
$w$ and adding the term $q/\rho$, we obtain
\begin{equation}
\label{hqrho}
w + \frac{q}{\rho} = \frac{\tilde E\gamma - G}{\gamma^2 - G} \enspace ,
\end{equation}
where the variable $\tilde E$ is given by
\begin{displaymath}
\tilde E = E + \frac{q}{\Gamma D}\enspace .
\end{displaymath}
Using
\begin{equation}
\label{ssquare}
S^2 = \eta^{ij}S_iS_j = \left(w + \frac{q}{\rho}\right)^2 \left(\gamma^2 -
1\right)
\end{equation}
and inserting expression (\ref{hqrho}), the relativistic $\gamma$-factor can
be determined explicitly from
\begin{equation}
\label{gamzeroeq}
0 = \left(S^2 - \tilde E^2\right) \gamma^4 + 2G\tilde E\gamma^3+\left(\tilde
E^2 - 2GS^2 - G^2\right)\gamma^2 - 2G\tilde E\gamma +
G^2\left(1+S^2\right)
\end{equation}
as the root of a polynomial of degree four. In the appendix we show that a
solution of equation (\ref{gamzeroeq}) exists and that it is unique for all
allowed values of $G$, $\tilde E$, and $S^2$. With the value of $\gamma$
known, one can calculate first the rest-mass density $\rho$ from equation
(\ref{densityvar}), then the thermodynamic pressure $p$ from equation
(\ref{hqrho}) and the equation of state (\ref{idealstateeq}), and finally the
velocity $\bar v^i$ from equation (\ref{momentumvar}) using $\eta^{ij}S_j=(w
+ q/\rho)\gamma\bar v^i$.
\section{\label{spheq}THE SPH EQUATIONS}
In this section we derive the SPH formalism for our set of relativistic
hydrodynamic equations (\ref{lagconti}), (\ref{lagmomentum}),
(\ref{lagenergy}), and (\ref{idealstateeq}). Since in the previous section
we have obtained the dynamic equations in an appropriate Lagrangian form, we
can proceed in a way that is completely analogous to the Newtonian case.
Before we derive our relativistic SPH equations, we give a brief introduction
into the standard SPH formalism, which was invented independently by Lucy
(1977) and Gingold \& Monaghan (1977). For a review of the SPH method see
Monaghan (1992).

SPH is a numerical method which discretizes the dynamic equations on a set of
nodes, called particles, moving with the fluid. The final discrete equations
are obtained in two separate steps. First, all hydrodynamic functions on
$\Sigma_t$ are smoothed over a certain volume with the help of a smoothing
kernel $W(|\mbox{\boldmath $r$} - \mbox{\boldmath$r$}^\prime |,h)$, i.e., for
a continuous function $f(\mbox{\boldmath$r$})$ one has
\begin{equation}
\label{intintpol}
f(\mbox{\boldmath$r$}) = \int f(\mbox{\boldmath$r$}')\,
W(|\mbox{\boldmath$r$}-\mbox{\boldmath$r$}^\prime|,h)\, d\mbox{\boldmath$r$}'
+ O(h^2)~,
\end{equation}
where $\mbox{\boldmath$r$}=\{x^i\}$ is the set of spatial coordinates $x^i$.
The kernel $W$ is a smooth (differentiable) function with compact support of
size $h$, the so called smoothing length, it is normalized to unity
\begin{displaymath}
\int W(|\mbox{\boldmath$r$}-\mbox{\boldmath$r$}^\prime|,h)\,
d\mbox{\boldmath$r$}' = 1~,
\end{displaymath}
and consequently, the smoothing error in equation (\ref{intintpol}) is
$O(h^2)$. The above integrals extend over a region on $\Sigma_t$ around
$\mbox{\boldmath$r$}$ that contains the support of $W$. An example for $W$ is
the cubic-spline kernel of Monaghan \& Lattanzio (1985), and in our
simulations we use this kernel throughout. The second step consists of an
approximate evaluation of the above integral (\ref{intintpol}) at the
particle positions $\mbox{\boldmath$r$}_a$
\begin{equation}
\label{sumintpol}
\int f(\mbox{\boldmath$r$}')\,
W(|\mbox{\boldmath$r$}_a-\mbox{\boldmath$r$}^\prime|,h)\,
d\mbox{\boldmath$r$}' \approx \sum_b \frac{f_b}{n_b}W_{ab} =:
\left<f\right>_a \enspace .
\end{equation}
Here, $a$ and $b$ label the particles which are distributed in space with
number density $n(\mbox{\boldmath$r$})$. We define $f_a =
f(\mbox{\boldmath$r$}_a)$ for any function $f$ and $W_{ab} =
W(|\mbox{\boldmath$r$}_a - \mbox{\boldmath$r$}_b|,h)$. The important point
is that, applying the smoothing and discretization operations
(\ref{intintpol}), (\ref{sumintpol}), it is possible to derive approximate
expressions for derivatives
\begin{equation}
\label{sphsumgrad}
\left<\mbox{\boldmath$\nabla$}f\right>_a\ = \sum_b \frac{f_b}{n_b}
\mbox{\boldmath$\nabla$}_aW_{ab}~~,~~~~ \left<\partial_if^i\right>_a\ =
\sum_b \frac{1}{n_b}{\mbox{\boldmath$f$}_b}
\mbox{\boldmath$\cdot\nabla$}_aW_{ab}
\enspace ,
\end{equation}
where $\mbox{\boldmath$f$}={\{f^i\}}$,
$\mbox{\boldmath$\nabla$}=\{\partial_i\}$, and
$\mbox{\boldmath$\nabla$}_aW_{ab}$ is the gradient of the kernel
$W(|\mbox{\boldmath$r$}_a - \mbox{\boldmath$r$}_b|,h)$ taken with respect to
the coordinates of particle $a$.

With the discretization method described above we are now in the position to
derive the SPH equations for our system of relativistic hydrodynamic
equations (\ref{lagconti}), (\ref{lagmomentum}), (\ref{lagenergy}), and
(\ref{idealstateeq}), and nothing has to be changed compared to the standard
(non-relativistic) SPH method. In the equations (\ref{sumintpol}) and
(\ref{sphsumgrad}) we replace the number density $n_a$ by the mass per
particle $m_a$, i.e., $n_a = D^*_a/m_a$. Dropping the angle brackets
$\left<~\right>$ from now on, equation (\ref{sumintpol}) leads to the SPH
representation for the relativistic rest-mass density
\begin{equation}
\label{sphdensity}
D^*_a = \sum_b m_b W_{ab} \enspace .
\end{equation}
From equation (\ref{lagconti}) and using 
\begin{displaymath}
D^*\mbox{\boldmath$\nabla\cdot v$} =
\mbox{\boldmath$\nabla\cdot$}(D^*\mbox{\boldmath$v$}) -
\mbox{\boldmath$v\cdot\nabla$}D^*\enspace ,
\end{displaymath}
we obtain the SPH form of the relativistic continuity equation
\begin{equation}
\label{sphconti}
\frac{d}{dt}D^*_a = - \sum_b m_b(\mbox{\boldmath$v$}_b -
\mbox{\boldmath$v$}_a)\mbox{\boldmath$\cdot\nabla$}_aW_{ab} \enspace .
\end{equation}
Here, $\mbox{\boldmath$v$}=\{v^i\}$ is just a shorthand notation for the set
of components $v^i$, whereas $\mbox{\boldmath$\nabla\cdot v$}$ stands for
$\partial_i v^i$ and is not the divergence of a vector field
$\mbox{\boldmath$v$}$. As a consequence, the total mass is conserved exactly.
Applying the Lagrangian time derivative to equation (\ref{sphdensity}) with
the smoothing length being constant in space and time, one can show that the
SPH expression of the relativistic rest-mass density $D^*$ in equation
(\ref{sphdensity}) automatically satisfies the relativistic continuity
equation. Thus, in SPH the relativistic rest-mass density $D^*$ can be
computed either from equation (\ref{sphdensity}) or equation
(\ref{sphconti}). If, however, the density $D = \gamma \rho$ is used instead
of $D^*$, a modification of the standard SPH method is required. Therefore,
Laguna et al. (1993) multiply the flat-space kernel $W(|\mbox{\boldmath$r$} -
\mbox{\boldmath$r$}'|,h)$ of Newtonian SPH with
$1/\sqrt{\eta(\mbox{\boldmath$r$}')}$. In a curved spacetime this factor
makes the kernel anisotropic, violates its translation invariance, and leads
to additional terms in the SPH approximation of derivatives. Applying the
relation $D^* = \sqrt{\eta}D$, the relativistic SPH formulation of Laguna
\etal (1993) can be cast into the standard SPH scheme which is considerably
simpler. We therefore suggest that there is no need to modify SPH for given
background spacetimes if the continuity equation is used in the conservative
form (\ref{lagconti}) without source terms.

In order to derive the SPH form of the relativistic momentum equation we
start from equation (\ref{lagmomentum}). Rewriting the pressure gradient as
\begin{displaymath}
\frac{1}{D^*}\mbox{\boldmath$\nabla$}\left[\sqrt{-g}(p+q)\right] =
\sqrt{-g}\left[\mbox{\boldmath$\nabla$}\left(\frac{p+q}{D^*}\right) +
\frac{p+q}{D^{*2}}\mbox{\boldmath$\nabla$}D^*\right] +
\frac{p+q}{D^*}\mbox{\boldmath$\nabla$}\sqrt{-g} \enspace ,
\end{displaymath}
we obtain
\begin{eqnarray}
\label{sphmomentumeq}
\frac{d}{dt}\mbox{\boldmath$S$}_a & = &-\sqrt{-g_a} \sum_b
m_b\left(\frac{p_a+q_a}{D^{*2}_a} +
\frac{p_b+q_b}{D^{*2}_b}\right)\mbox{\boldmath$\nabla$}_aW_{ab}\nonumber\\
&&{}-\frac{\sqrt{-g_a}}{D^*_a}\left[(p_a+q_a)\mbox{\boldmath$\nabla$}_a
\left(\ln{\sqrt{-g}}\right)_a-\frac{1}{2}T^{\alpha\beta}_a
\mbox{\boldmath$\nabla$}_a\left(g_{\alpha\beta}\right)_a\right]
\enspace ,
\end{eqnarray}
where $\mbox{\boldmath$S$}_a={\{S_i\}}_a$, and the metric gradients
$\mbox{\boldmath$\nabla$}\ln\sqrt{-g}$ and
$\mbox{\boldmath$\nabla$}g_{\alpha\beta}$ can be calculated analytically.
Thus, only the pressure gradient term in equation (\ref{sphmomentumeq}) has
to be smoothed, and it has been symmetrized to conserve linear and angular
momentum in SPH. Next, we proceed with the relativistic energy equation
(\ref{lagenergy}). In the expression
\begin{displaymath}
\frac{1}{D^*} \mbox{\boldmath$\nabla\cdot$} 
\left[\sqrt{-g}(p+q)\mbox{\boldmath$v$}\right] = \frac{\sqrt{-g}}{D^*}
\mbox{\boldmath$\nabla\cdot$} \left[(p+q)\mbox{\boldmath$v$}\right] +
\frac{p+q}{D^*} \mbox{\boldmath$v\cdot\nabla$}\sqrt{-g}
\end{displaymath}
we replace the first term on the right hand side by
\begin{eqnarray*}
\frac{1}{D^*} \mbox{\boldmath$\nabla\cdot$} \left[(p+q)
\mbox{\boldmath$v$}\right] & = & \frac{1}{D^*} \mbox{\boldmath$v\cdot\nabla$}
(p+q) + \frac{p+q}{D^*} \mbox{\boldmath$\nabla\cdot v$}
\nonumber\\
& = &\mbox{\boldmath$v\cdot$} \left[ \mbox{\boldmath$\nabla$}
\left(\frac{p+q}{D^*} \right) + \frac{p+q}{D^{*^2}} \mbox{\boldmath$\nabla$}
D^* \right] \nonumber\\
&&{}+\frac{1}{2} \left[ \mbox{\boldmath$\nabla\cdot$} \left(\frac{p+q}{D^*}
\mbox{\boldmath$v$} \right) - \mbox{\boldmath$v\cdot\nabla$}
\left(\frac{p+q}{D^*}\right) \right. \nonumber\\
&&\left.{}+\frac{p+q}{D^{*2}} \left[ \mbox{\boldmath$\nabla\cdot$}
(D^* \mbox{\boldmath$v$}) - \mbox{\boldmath$v\cdot\nabla$} D^* \right]
\right] \enspace .
\end{eqnarray*}
With this combination of terms we obtain the SPH representation of the
relativistic energy equation
\begin{eqnarray}
\label{sphenergyeq}
%
\frac{d}{dt} {\left[ \alpha E - \beta^i S_i \right]}_a & = &
-\frac{\sqrt{-g_a}}{2} \sum_b m_b\left(\frac{p_a+q_a}{D^{*2}_a} +
\frac{p_b+q_b}{D^{*2}_b}\right) (\mbox{\boldmath$v$}_a +
\mbox{\boldmath$v$}_b) \mbox{\boldmath$\cdot\nabla$}_a W_{ab} \nonumber\\
&&{}-\frac{\sqrt{-g_a}}{D^*_a} \left[ (p_a+q_a) \mbox{\boldmath$v$}_a 
{\mbox{\boldmath$\cdot\nabla$}_a \left(\ln{\sqrt{-g}}\right)}_a + \frac{1}{2}
T^{\alpha\beta}_a {\left(g_{\alpha\beta,t}\right)}_a \right]
\enspace .
\end{eqnarray}
Note that equation (\ref{sphenergyeq}) and the momentum equation
(\ref{sphmomentumeq}) contain identical symmetric factors in front of the
kernel gradients. As outlined in section \ref{hydroeq}, this form of the
relativistic energy equation is well suited for SPH because it contains no
time derivatives of hydrodynamic variables on its right hand side. Again,
there is no need to smooth the derivatives of the metric
$\mbox{\boldmath$\nabla$}\ln\sqrt{-g}$ and $g_{\alpha\beta,t}$. The last
equation that needs to be considered is the equation of state
(\ref{idealstateeq}), and its SPH formulation reads
\begin{equation}
\label{sphstateeq}
p_a = (\Gamma - 1)\rho_a\varepsilon_a \enspace .
\end{equation}
This completes our set of relativistic SPH equations (\ref{sphdensity}),
(\ref{sphmomentumeq}), (\ref{sphenergyeq}), and (\ref{sphstateeq}) for
computing general relativistic fluid flows in given arbitrary background
spacetimes.

We now focus our attention to the implementation of an artificial viscosity
term which is necessary to handle shock fronts. For the use of an artificial
viscosity in the standard SPH method see Monaghan \& Gingold (1983). In our
simulations, we have used the following artificial viscous pressure
\begin{equation}
\label{artvisc}
q_a = \left\{\begin{array}{ll} \rho_a w_a \left[ - \tilde\alpha c_a h_a
{\left( \mbox{\boldmath$\nabla\cdot v$} \right)}_a + \tilde\beta h_a^2
{\left( \mbox{\boldmath$\nabla\cdot v$} \right)}_a^2 \right] &
\mbox{if}\enspace {\left( \mbox{\boldmath$\nabla\cdot v$} \right)}_a < 0
\\ 0 & \mbox{otherwise}\end{array}\right.
\end{equation}
with
\begin{equation}
\label{sphdivv}
{\left( \mbox{\boldmath$\nabla\cdot v$} \right)}_a \approx
\frac{\mbox{\boldmath$v$}_{ab} \mbox{\boldmath$\cdot r$}_{ab}}
{|\mbox{\boldmath$r$}_{ab}|^2 + \tilde\varepsilon \bar h^2_{ab}} \enspace ,
\end{equation}
where $c_a=\sqrt{\Gamma p_a/(\rho_aw_a)}$ is the relativistic sound velocity
measured in the rest frame of the fluid, $\tilde\alpha$, $\tilde\beta$ and
$\tilde\varepsilon$ are numerical parameters, $\mbox{\boldmath$v$}_{ab}$,
$\mbox{\boldmath$r$}_{ab}$ stand for the differences
$\mbox{\boldmath$v$}_{ab} = \mbox{\boldmath$v$}_a - \mbox{\boldmath$v$}_b$,
$\mbox{\boldmath$r$}_{ab} = \mbox{\boldmath$r$}_a - \mbox{\boldmath$r$}_b$,
and $\bar h_{ab}=(h_a + h_b)/2$ is the mean value of the smoothing lengths of
particles $a$ and $b$. Without the enthalpy $w_a$, the expression
(\ref{artvisc}) for $q$ is almost equivalent to the standard SPH form of the
artificial viscosity invented by Monaghan \& Gingold (1983). Including $w_a$
into equation (\ref{artvisc}), the parameters $\tilde\alpha$ and
$\tilde\beta$ can be chosen to be of order unity even for shocks with
ultra-relativistic values of $\gamma$. The $\tilde\alpha$-term, which is
linear in the velocity differences, is similar to a physical shear and bulk
viscosity, and the quadratic $\tilde\beta$-term is the standard von
Neumann-Richtmyer (1950) artificial viscosity used in finite difference
methods for handling high Mach-number shocks. According to Monaghan \&
Gingold (1983), the un-smoothed representation of the velocity divergence in
equation (\ref{sphdivv}) acts more directly on the relative motion of
particle pairs and leads to a damping of irregular oscillations in shock
transitions. In equation (\ref{sphdivv}), the parameter $\tilde\varepsilon$
has been introduced to avoid singularities, and a typical value is
$\tilde\varepsilon=0.1$.

Since the expression (\ref{artvisc}) for the artificial viscosity is not
Lorentz invariant, we also performed calculations with a relativistically
covariant formulation of the artificial viscous pressure
\begin{equation}
\label{artviscrel}
q_a = \left\{\begin{array}{ll} \rho_a\left[-\tilde\alpha c_ah_a\theta_a +
\tilde\beta h_a^2\theta_a^2\right] & \mbox{if}\enspace \theta_a<0\\ 0 &
\mbox{otherwise}\end{array}\right.
\end{equation}
where
\begin{displaymath}
\theta_a = {\left({u^\mu}_{;\mu}\right)}_a = \frac{1}{\alpha_a} \left[
\frac{d\gamma_a}{dt} + \gamma_a \left( {\left( \mbox{\boldmath$\nabla\cdot
v$} \right)}_a + \frac{d}{dt} {\left(\ln\sqrt{\eta}\right)}_a \right) \right]
\enspace .
\end{displaymath}
However, this expression contains a time derivative of the $\gamma$-factor
which destroys the explicit nature of the Lagrangian form of the hydrodynamic
equations. One possibility to circumvent this problem is to take the backward
time difference approximation $[\gamma_a(t) - \gamma_a(t-\Delta t)]/\Delta t$
with time step $\Delta t$ for the time derivative $d\gamma_a/dt$. Since the
non-covariant form (\ref{artvisc}) turned out to be quite appropriate for the
resolution of shock structures, we used this approach in all our simulations
and did not pursue the covariant relation (\ref{artviscrel}).

To improve the local resolution of SPH, we allow the smoothing length $h$ to
vary in space according to the relativistic rest-mass density $D^*$ via
\begin{equation}
\label{varhsml}
h_a = \left(h_0\right)_a\left[\frac{\left(D^*_0\right)_a}{D^*_a}\right]^{1/d}
\enspace ,
\end{equation}
where $d$ denotes the dimension of the spatial slices $\Sigma_t$. For
particles of equal mass the scaling law (\ref{varhsml}) indicates that the
number of particles within the support of the kernel $W$ is approximately
kept constant in time. Thus, in regions where the gas is compressed, the
smoothing length is increased while it is decreased in rarefaction zones.
Since the smoothing length is now a function of the density $D^*$, equation
(\ref{sphdensity}) (or eq.~[\ref{sphconti}]) is now a nonlinear implicit
relation for the density. We solve this approximately by inserting $D^*_a$
from the previous time step into equation (\ref{varhsml}) to obtain an
estimate for $h_a$. Since the smoothing length is now a function of position
and time, the SPH form of the continuity equation contains additional terms
(Monaghan 1992). However, no such terms appear if the relativistic rest-mass
density $D^*_a$ is calculated from the computationally simpler equation
(\ref{sphdensity}).
\section{\label{codetests}NUMERICAL TESTS}
Although we have developed a fully three-dimensional general relativistic SPH
code, we restrict ourselves in this paper to the standard analytic test bed
of one-dimensional special relativistic shock problems: the shock tube and
the wall shock. For each simulation we show four diagrams of the numerical
results together with the analytic solution for the fluid variables
3-velocity $v:=\bar v^1=v^1$, rest-mass density $\rho$, thermodynamic
pressure $p$, and specific internal energy $\varepsilon$. These variables are
functions of the coordinate $x:=x^1 \in [0,100]$. The quality of the
simulations is measured in terms of the relative error with respect to the
analytic solution, i.e., for each hydrodynamic function $f$ an error $\Delta
f$ is calculated from
\begin{displaymath}
\Delta f = \frac{1}{Nf^0_{\rm max}}\sum^N_{b=1}|f_b - f^0_b| \enspace ,
\end{displaymath}
where the sum is over all $N$ particles and $f^0$ stands for the analytic
solution which has the maximum value $f^0_{\rm max}$.

\subsubsection*{a) Shock Tube Tests}

First we consider the shock tube problem, where initially a fluid at rest is
divided by a diaphragm into two regions of different densities and internal
energies. When the diaphragm is removed, a rarefaction wave travels into the
warm and dense medium and a compression wave into the cold and lower density
fluid. Between the two media a so called contact discontinuity is present.
For the analytic solution of the relativistic shock tube problem we refer to
Taub (1948), McKee \& Colgate (1973), Hawley \etal (1984a), and Marti \&
M\"uller (1994).

\begin{figure}[t]
\begin{center}
\begin{minipage}[t]{\figuresize}
\epsfxsize=\boxsize
\epsfbox{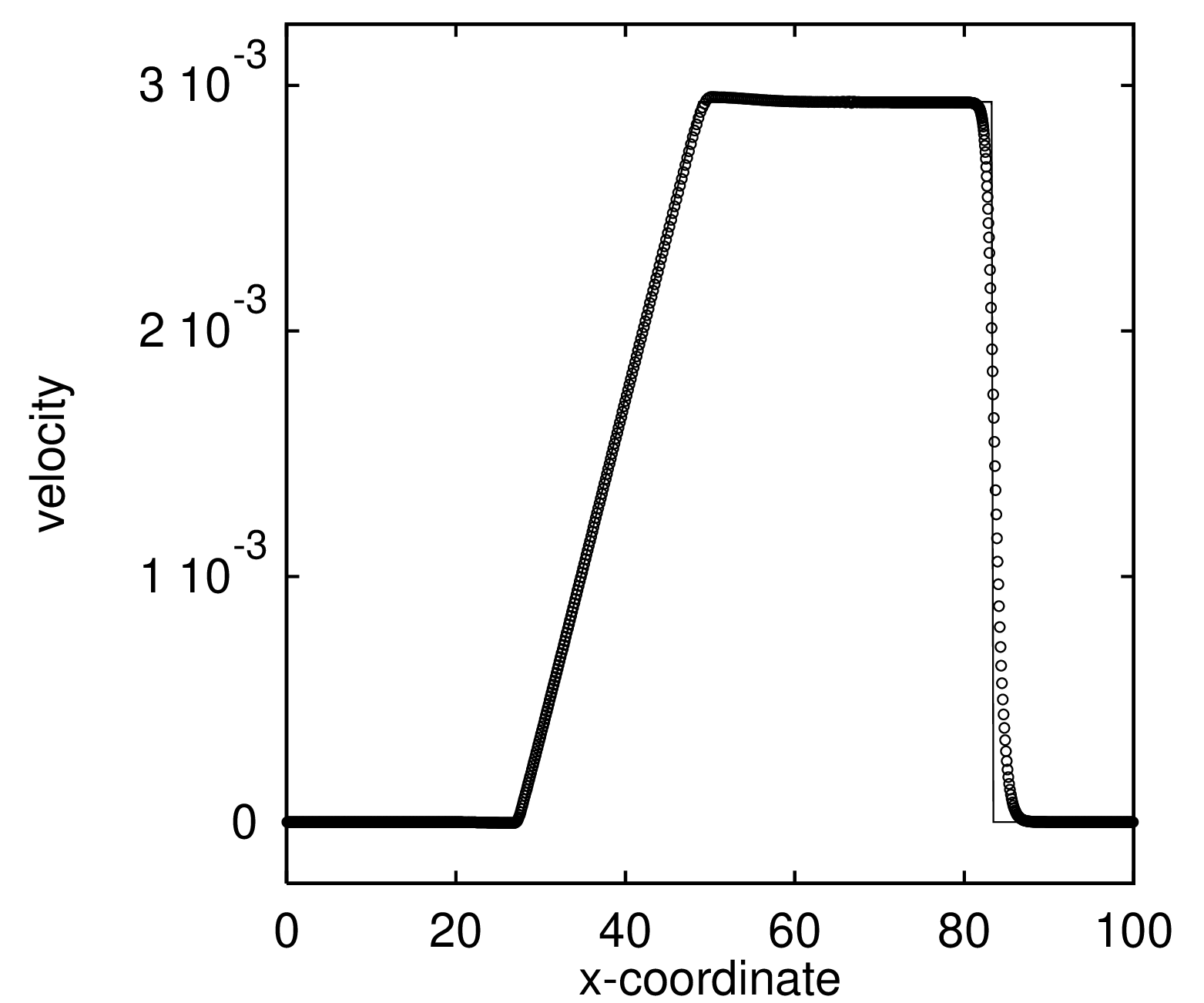}
\end{minipage}
\hspace{15mm}
\begin{minipage}[t]{\figuresize}
\epsfxsize=\boxsize
\epsfbox{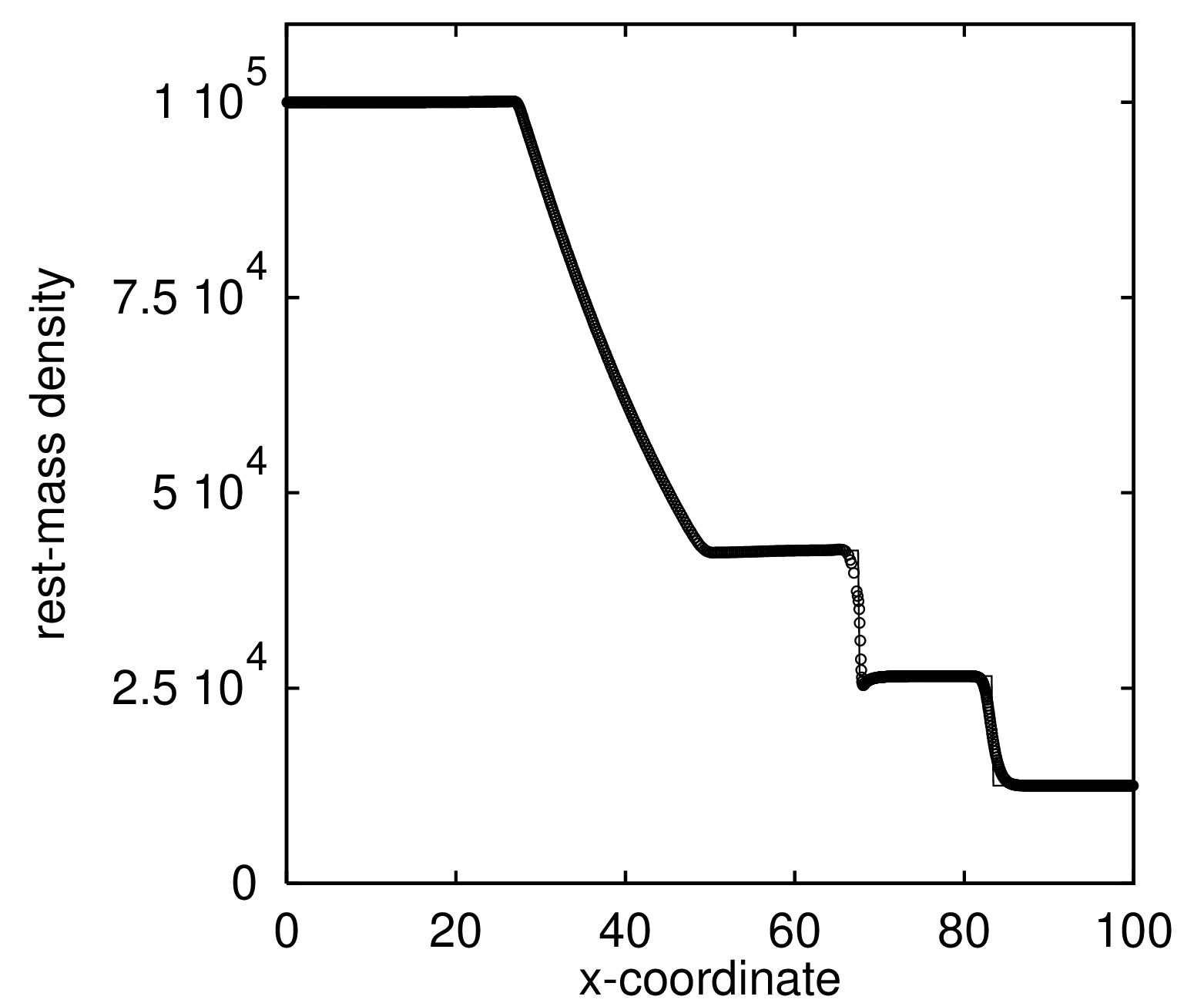}
\end{minipage}
\\
\vspace{10mm}
\begin{minipage}[t]{\figuresize}
\epsfxsize=\boxsize
\epsfbox{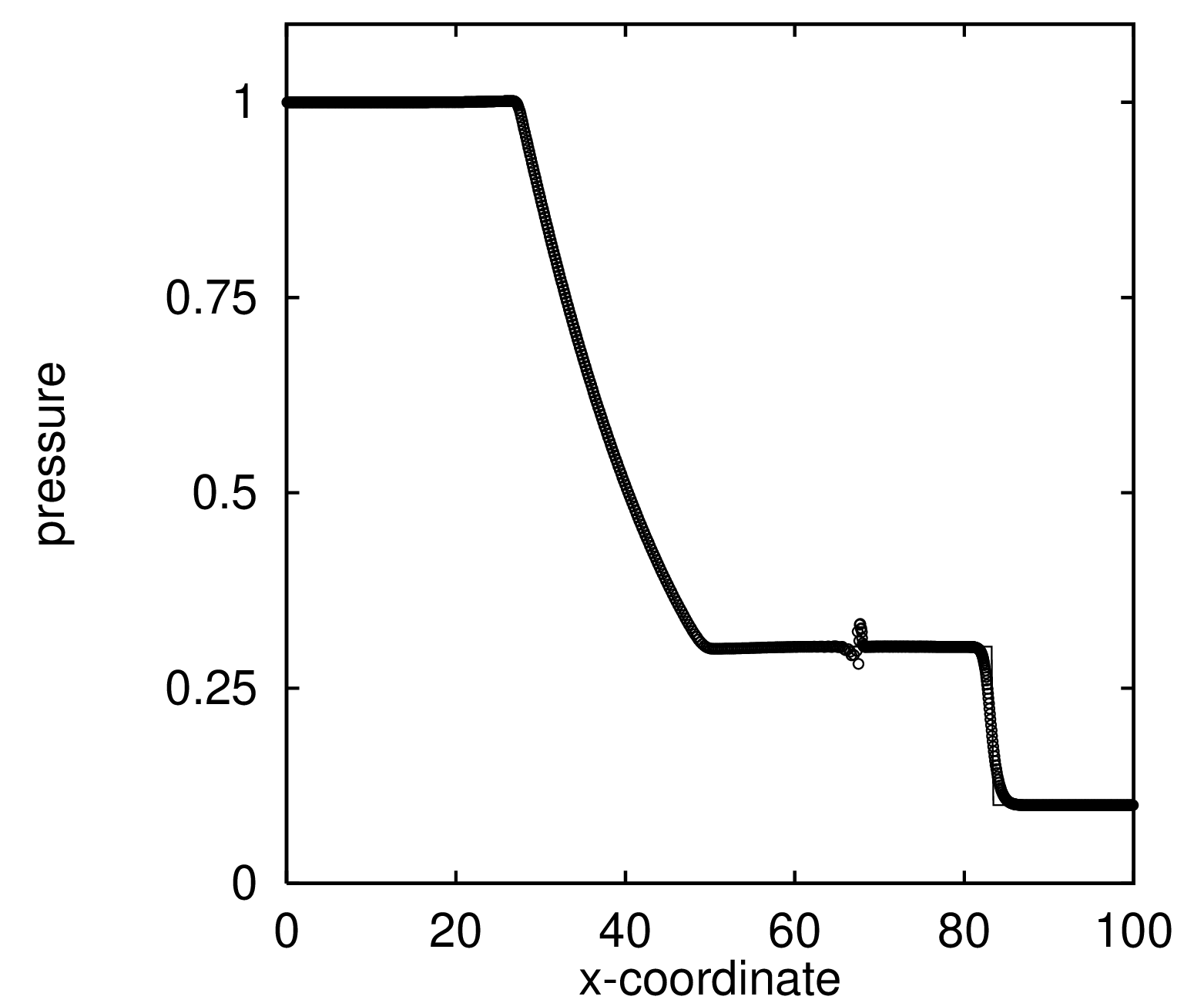}
\end{minipage}
\hspace{15mm}
\begin{minipage}[t]{\figuresize}
\epsfxsize=\boxsize
\epsfbox{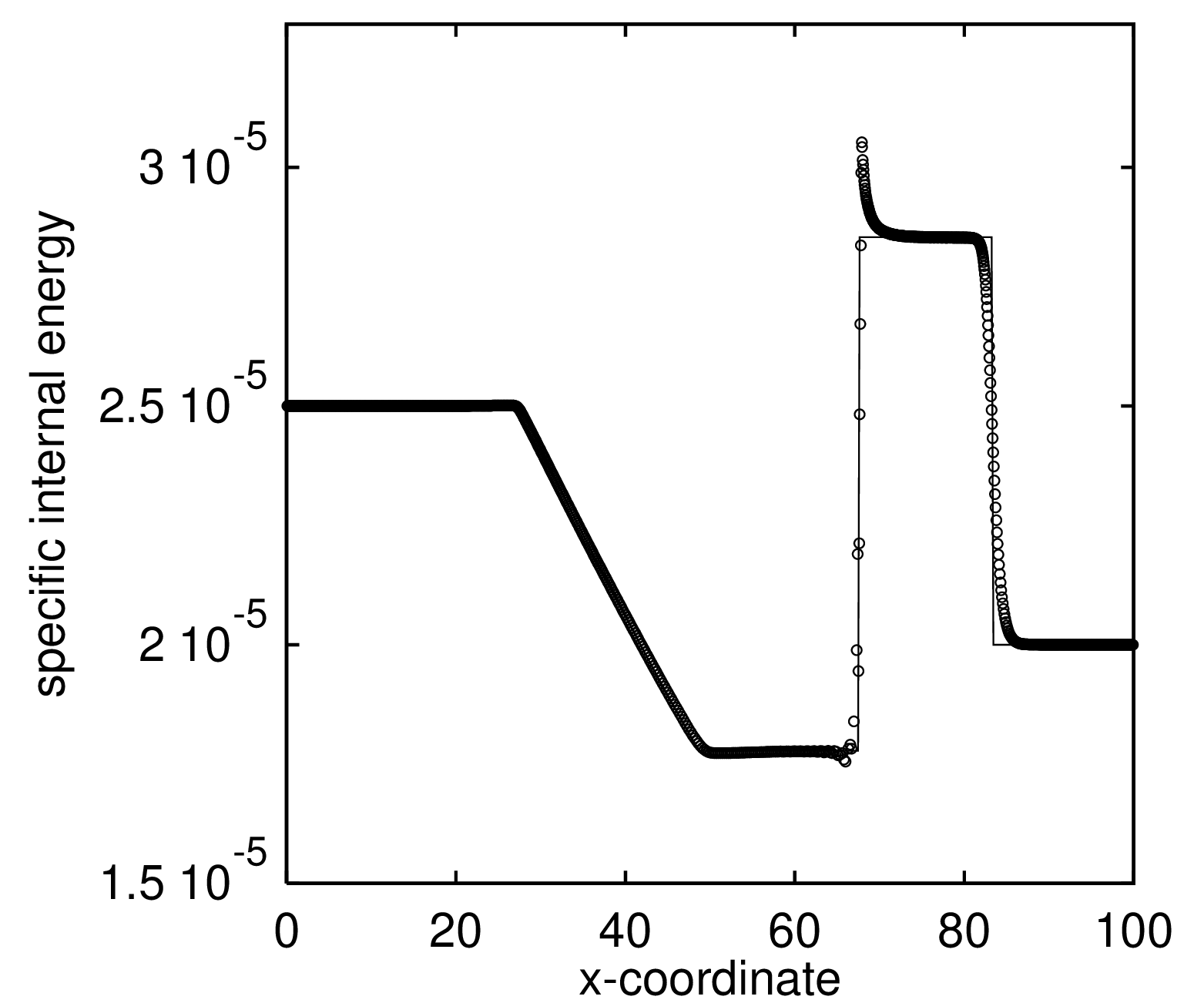}
\end{minipage}
\end{center}
\caption{\label{fig1} Numerical result of an SPH calculation with $1000$
particles (open circles) and analytic solution (solid line) for the
non-relativistic shock tube problem (note that the units are arbitrary except
for the velocity which is measured in units of $c$). The intermediate
pressure, velocity, and relativistic $\gamma$-factor are $p_m=0.303$,
$v_m=2.93\times10^{-3}$, and $\gamma_m=1.000004$, the positions of the shock,
the contact discontinuity, and the head and tail of the rarefaction wave are
$x_s=83.2$, $x_c=67.6$, $x_{h}=27.6$, and $x_{t}=48.7$, respectively, and the
velocity and relativistic $\gamma$-factor of the shock are
$v_s=5.54\times10^{-3}$ and $\gamma_s=1.000015$.}
\end{figure}

As any relativistic hydro code has to be tested for the Newtonian limit, we
start our series of shock tube simulations with a non-relativistic test case.
Figure \ref{fig1} shows the corresponding numerical and analytic solution
for a gas with $\Gamma=1.4$ at time $t=6000$ with the initial conditions
$\rho=10^5$, $\varepsilon=2.5\times10^{-5}$ for $x < 50$, and
$\rho=0.125\times10^5$, $\varepsilon=2\times10^{-5}$ for $x > 50$ (note that
the units are arbitrary except for the velocity which is measured in units of
$c$). The numerical calculation was performed with 1000 particles, initial
smoothing length $h=0.6$ ($\approx10$ interactions per particle), and
artificial viscosity parameters $\tilde\alpha=1$ and $\tilde\beta=2$. In the
initial distribution, particles were placed on a uniform grid, and the
particles at $x>50$ have a mass ten times smaller than the particles at
$x<50$. In the calculation of Figure \ref{fig1} the largest relative error
occurs in the fluid velocity where $\Delta v=1.1\%$.

\begin{figure}[t]
\begin{center}
\begin{minipage}[t]{\figuresize}
\epsfxsize=\boxsize
\epsfbox{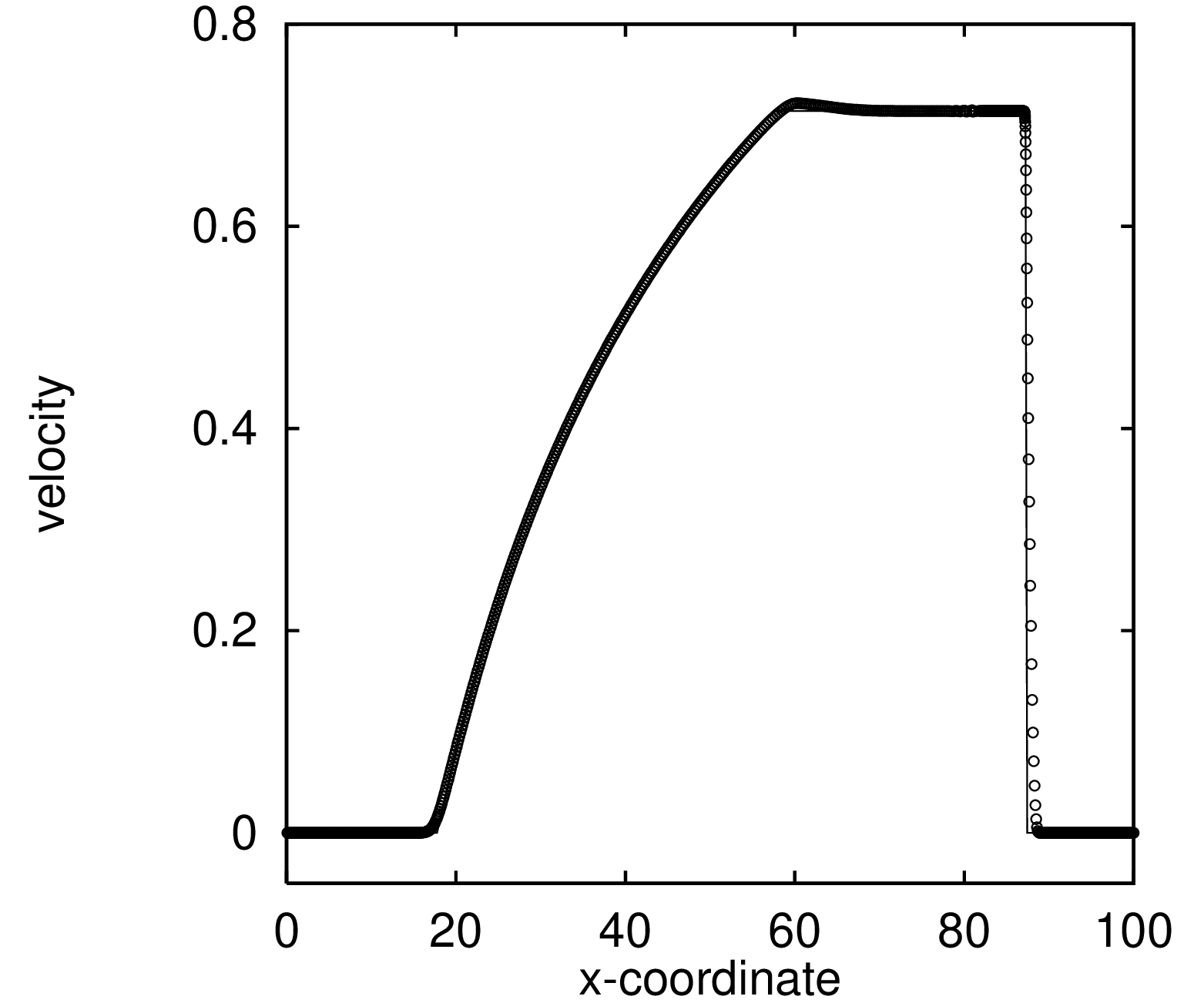}
\end{minipage}
\hspace{15mm}
\begin{minipage}[t]{\figuresize}
\epsfxsize=\boxsize
\epsfbox{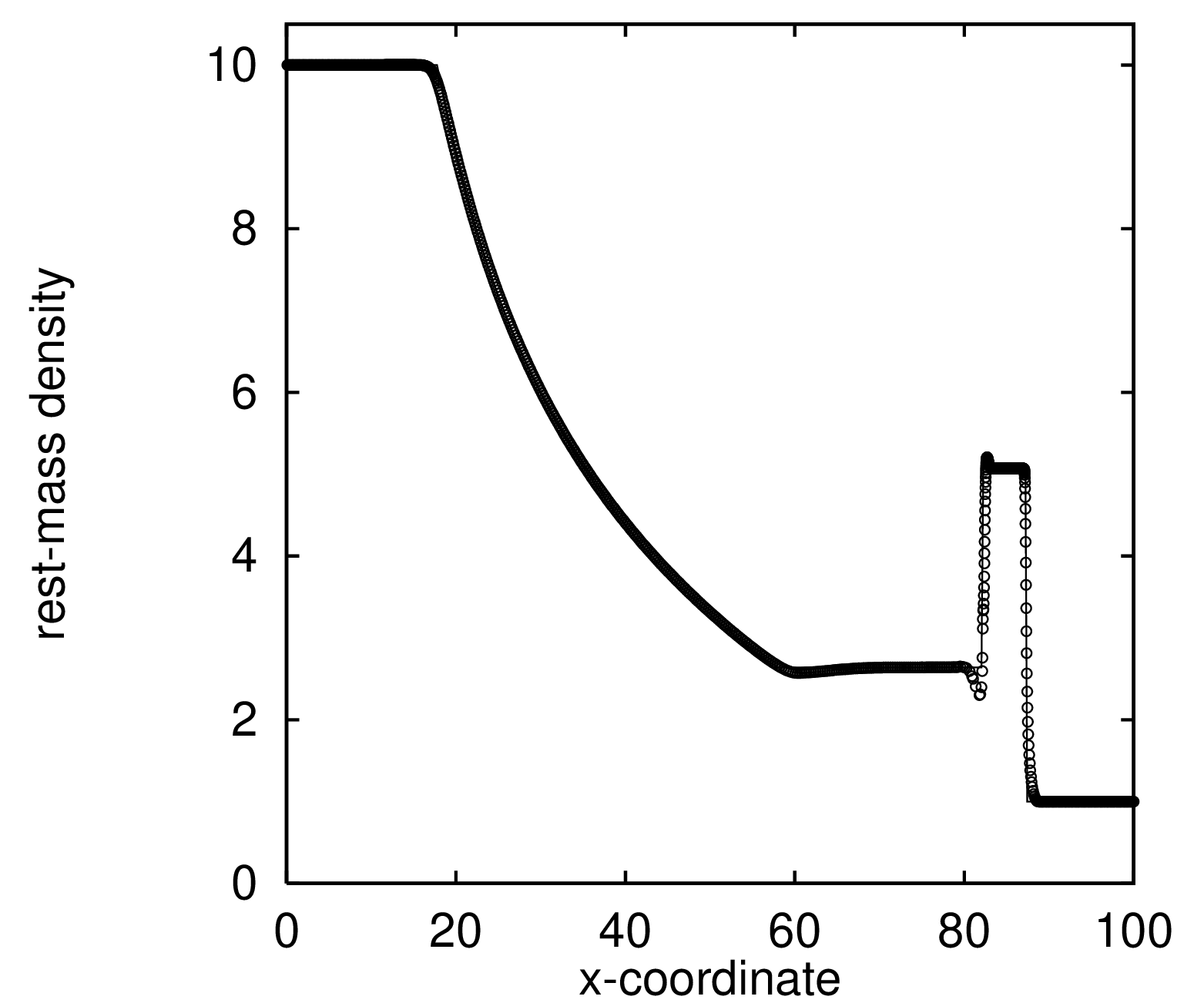}
\end{minipage}
\\
\vspace{10mm}
\begin{minipage}[t]{\figuresize}
\epsfxsize=\boxsize
\epsfbox{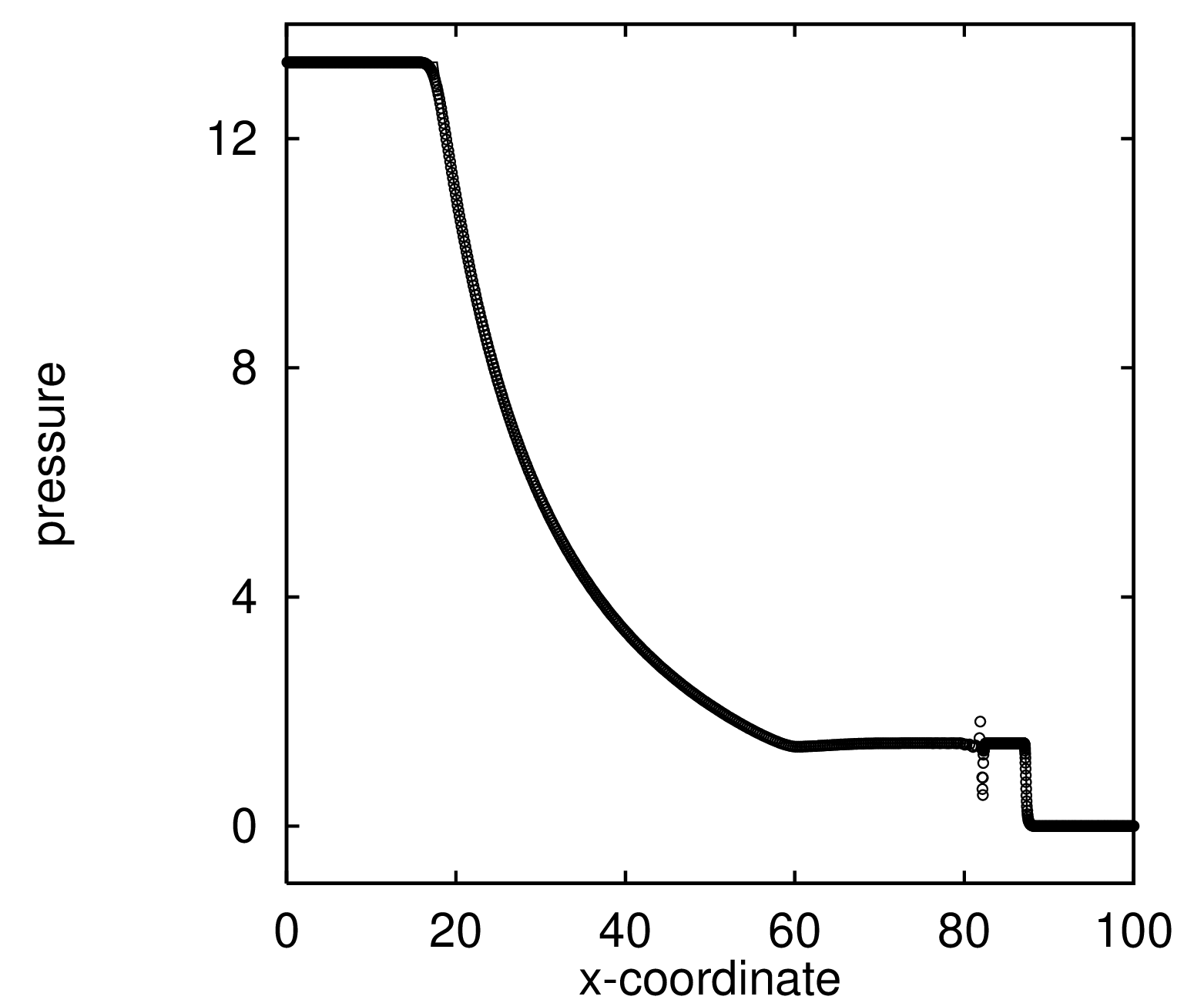}
\end{minipage}
\hspace{15mm}
\begin{minipage}[t]{\figuresize}
\epsfxsize=\boxsize
\epsfbox{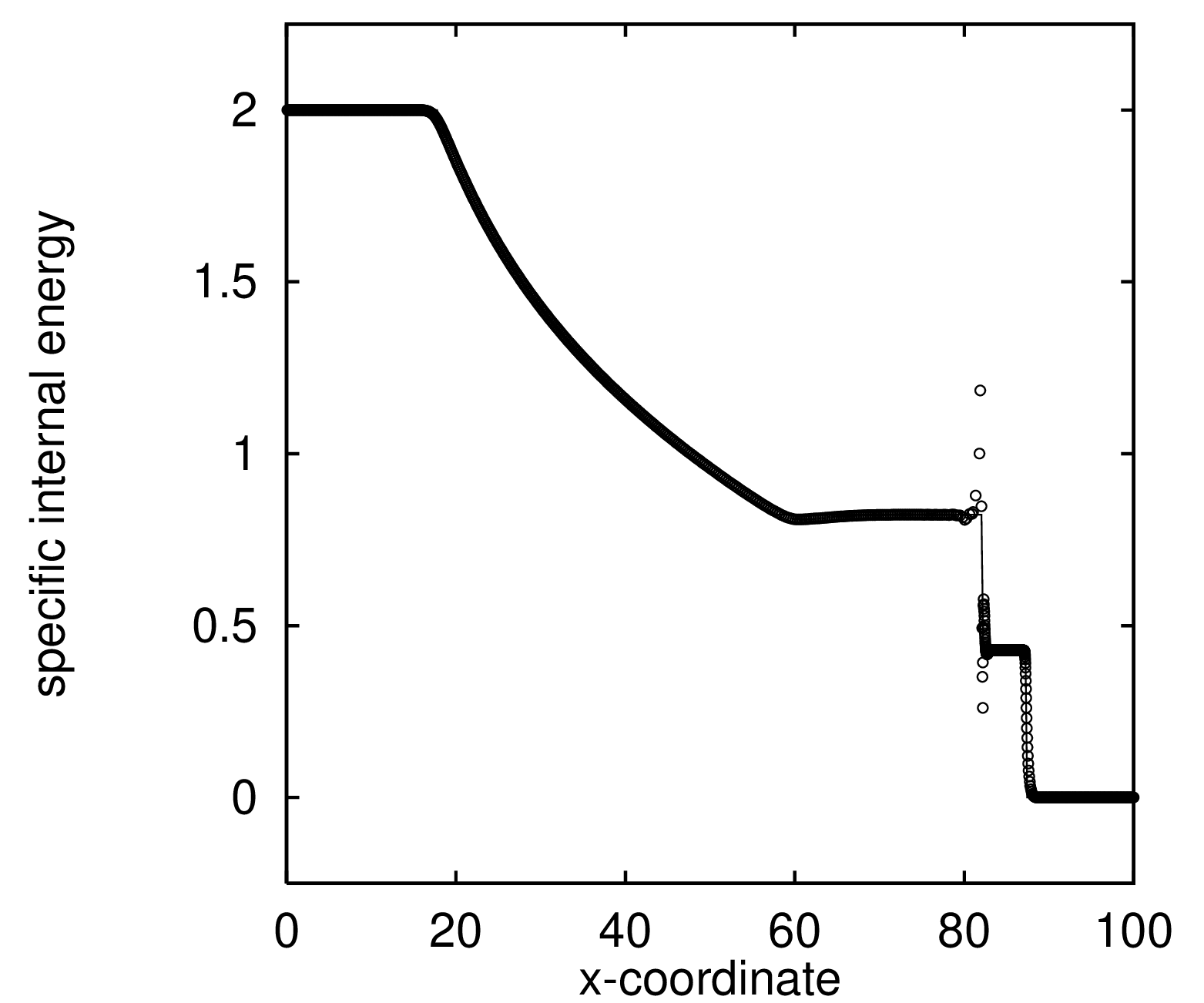}
\end{minipage}
\end{center}
\caption{\label{fig2} Numerical result of an SPH calculation with $1000$
particles (open circles) and analytic solution (solid line) for the
relativistic shock tube problem. The intermediate pressure, velocity, and
relativistic $\gamma$-factor are $p_m=1.45$, $v_m=0.714$, and $\gamma_m=1.4$,
the positions of the shock, the contact discontinuity, and the head and tail
of the rarefaction wave are $x_s=87.3$, $x_c=82.1$, $x_{h}=17.8$, and
$x_{t}=57.5$, respectively, and the velocity and relativistic $\gamma$-factor
of the shock are $v_s=0.828$ and $\gamma_s=1.8$.}
\end{figure}

\newpage

\begin{figure}[t]
\begin{center}
\begin{minipage}[t]{\figuresize}
\epsfxsize=\boxsize
\epsfbox{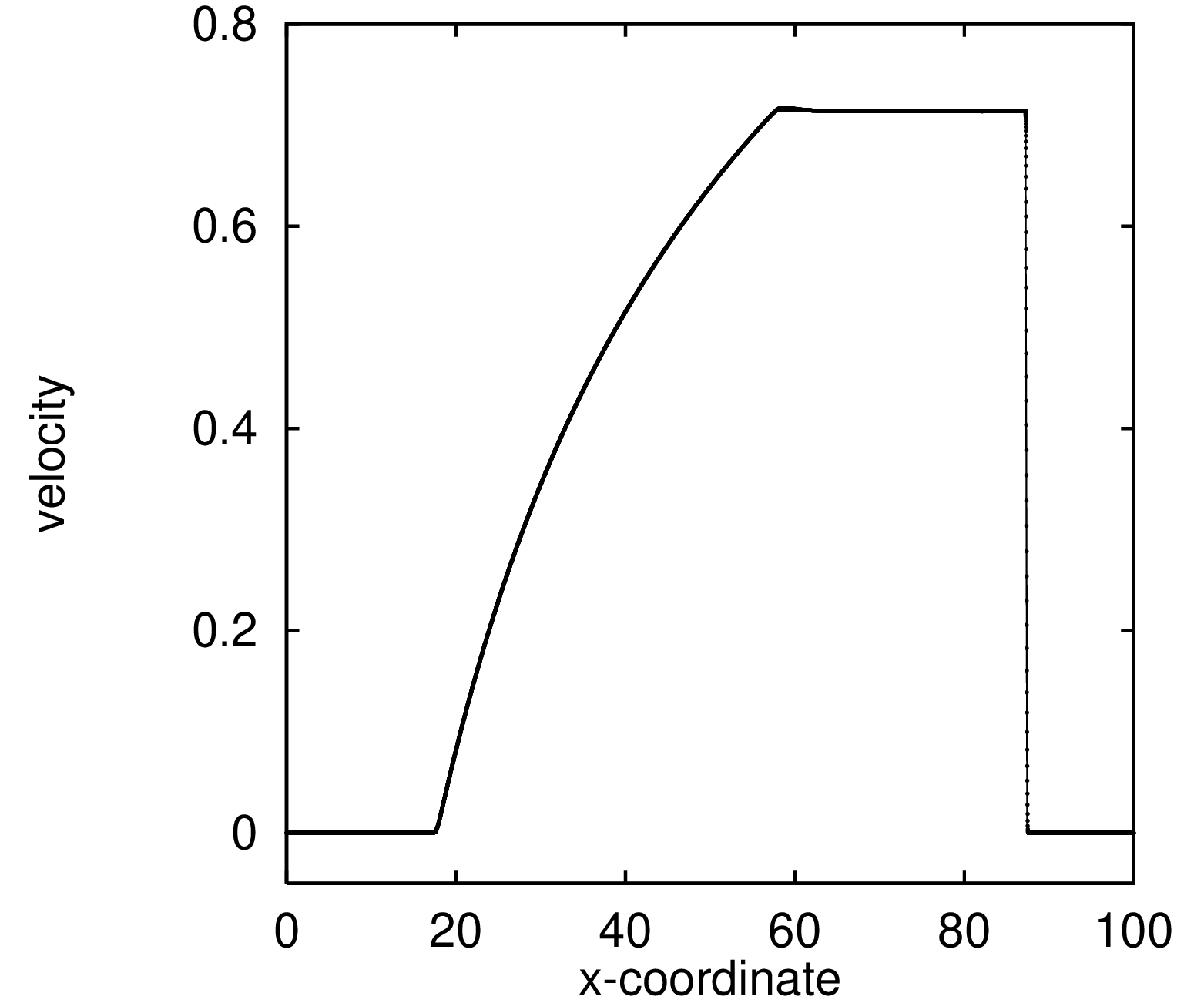}
\end{minipage}
\hspace{15mm}
\begin{minipage}[t]{\figuresize}
\epsfxsize=\boxsize
\epsfbox{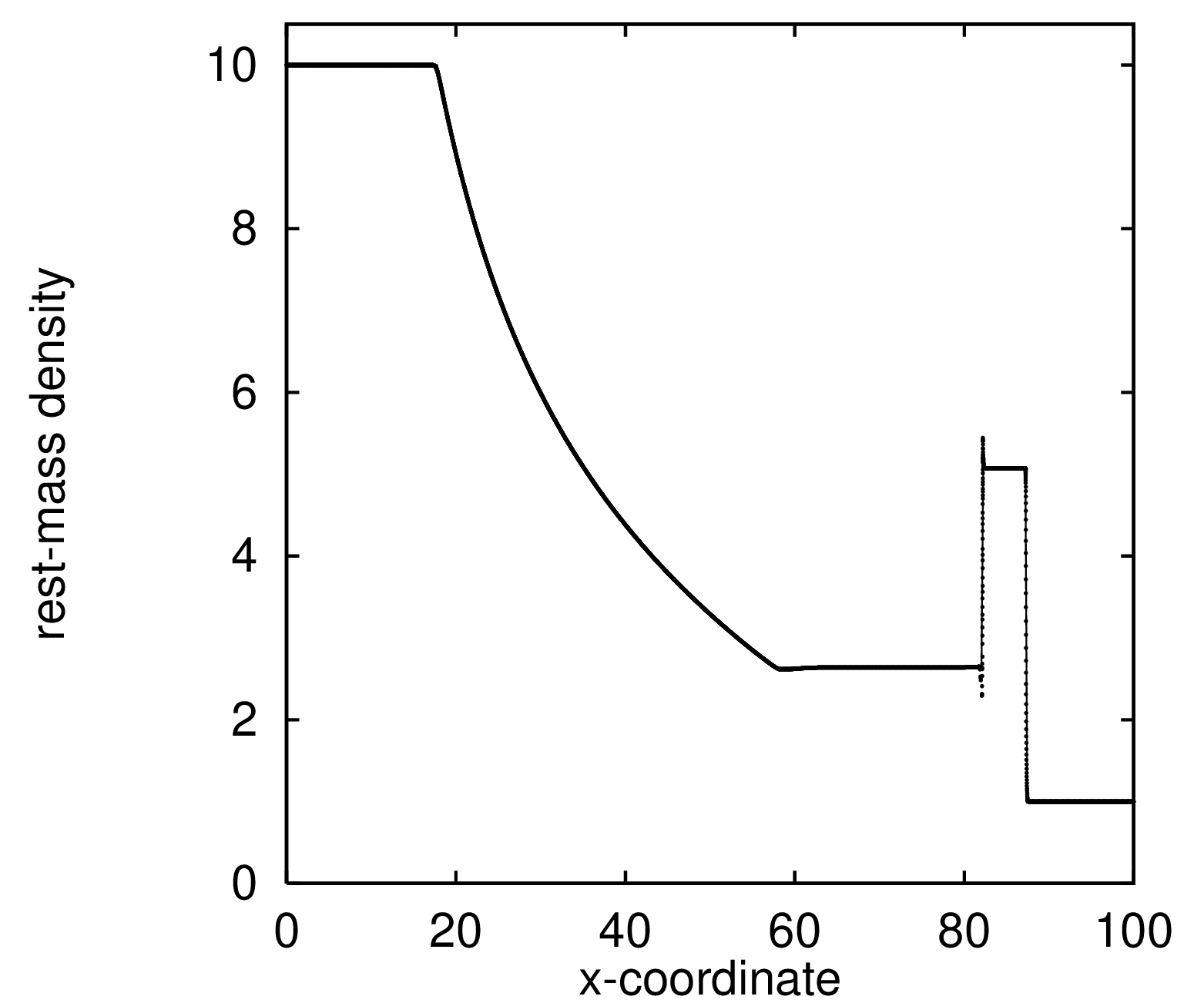}
\end{minipage}
\\
\vspace{10mm}
\begin{minipage}[t]{\figuresize}
\epsfxsize=\boxsize
\epsfbox{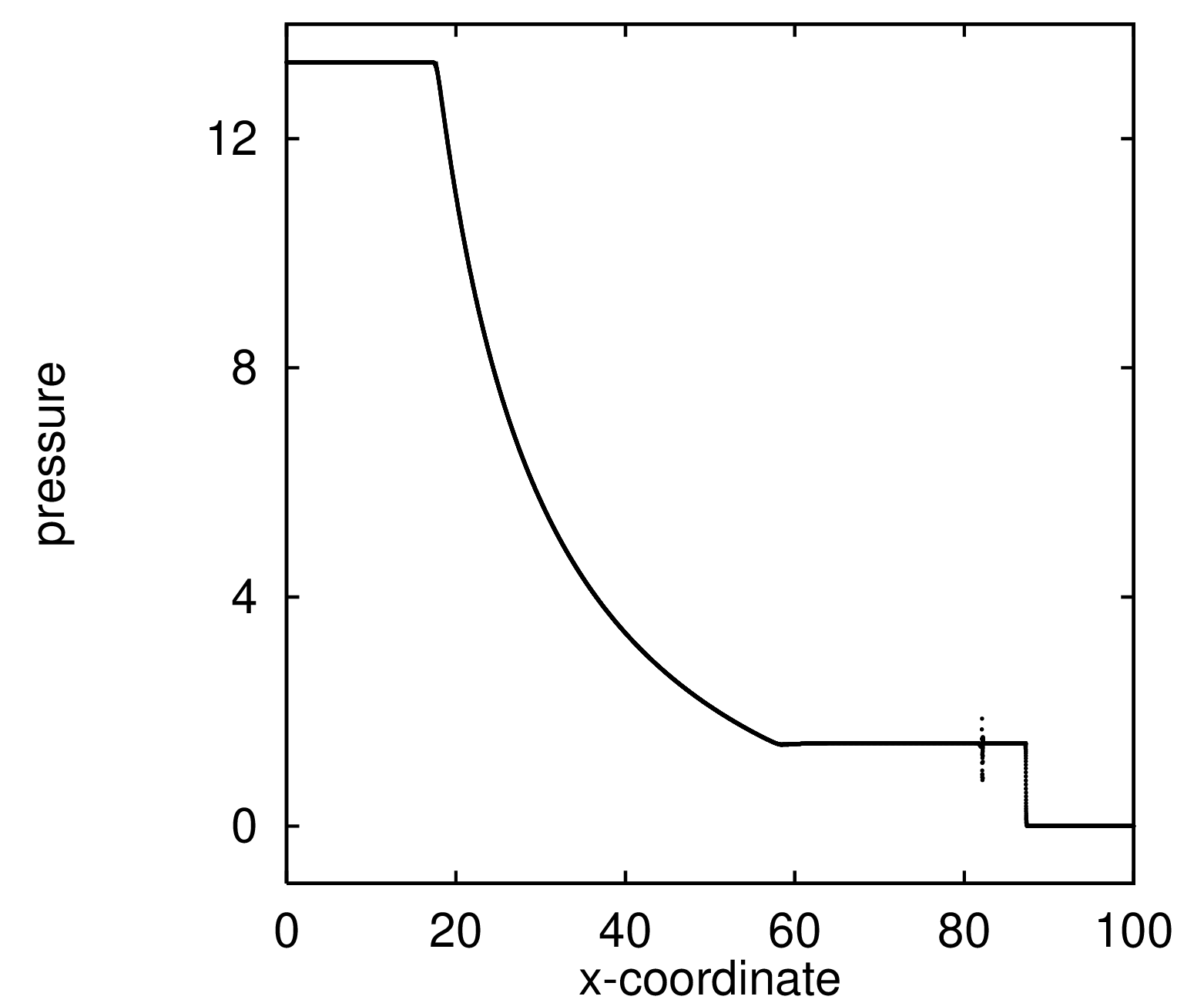}
\end{minipage}
\hspace{15mm}
\begin{minipage}[t]{\figuresize}
\epsfxsize=\boxsize
\epsfbox{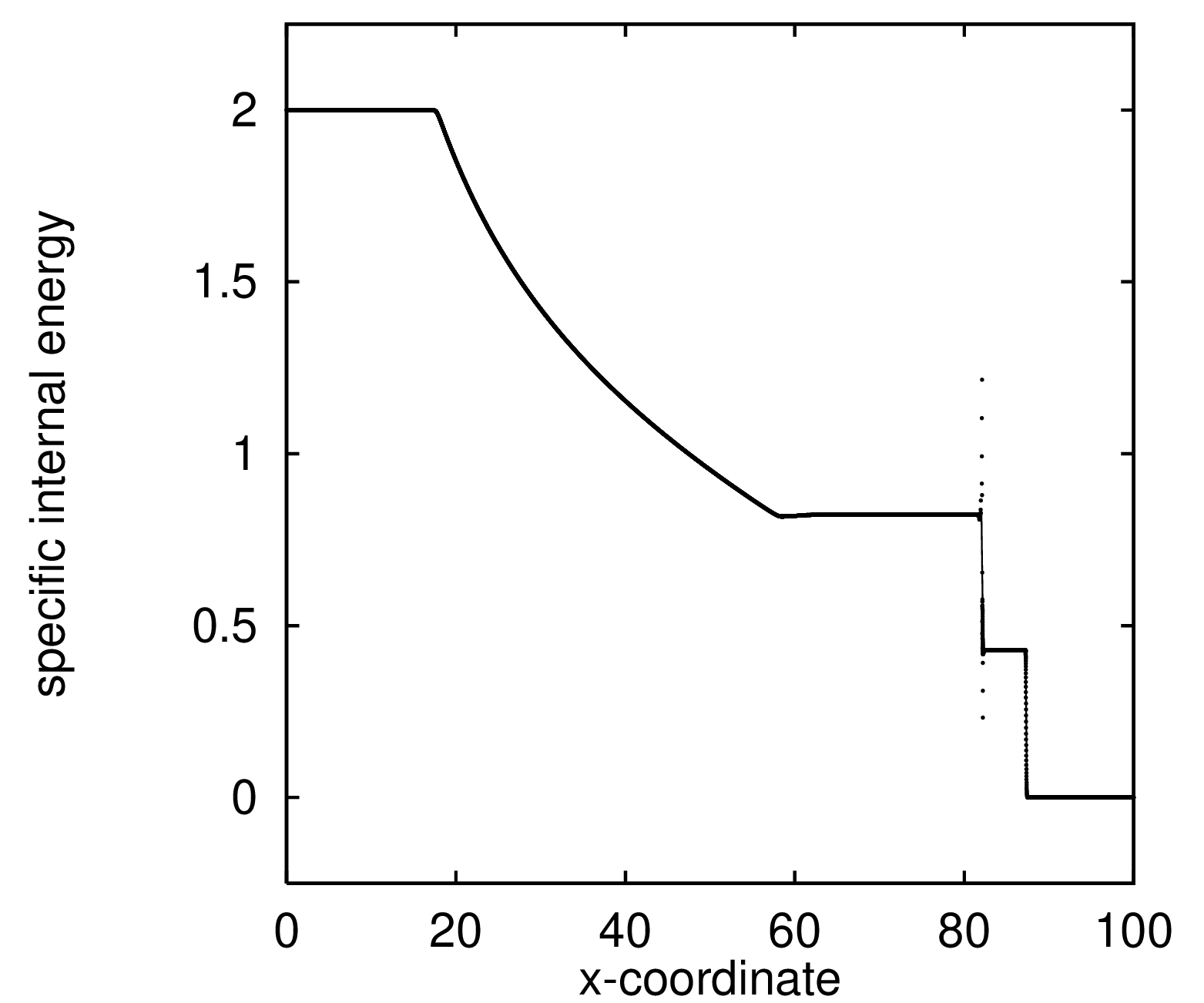}
\end{minipage}
\end{center}
\caption{\label{fig3} Numerical result of an SPH calculation with $10000$
particles (points) and analytic solution (solid line) for the relativistic
shock tube problem. The intermediate pressure, velocity, and relativistic
$\gamma$-factor are $p_m=1.45$, $v_m=0.714$, and $\gamma_m=1.4$, the
positions of the shock, the contact discontinuity, and the head and tail of
the rarefaction wave are $x_s=87.3$, $x_c=82.1$, $x_{h}=17.8$, and
$x_{t}=57.5$, respectively, and the velocity and relativistic $\gamma$-factor
of the shock are $v_s=0.828$ and $\gamma_s=1.8$.}
\end{figure}

Next, a mildly relativistic shock tube is investigated with initial
conditions $\rho=10$, $\varepsilon=2$ ($x<50$), and $\rho=1$,
$\varepsilon=10^{-6}$ ($x>50$). Figure \ref{fig2} shows the numerical result
and the analytic solution for a gas with $\Gamma=5/3$ at time $t=45$. We used
the same numerical parameters as in the non-relativistic shock tube problem,
i.e., 1000 particles, $h=0.6$, $\tilde\alpha=1$, and $\tilde\beta=2$. The
largest relative error is $\Delta v=1.0\%$. As can be seen in Figure
\ref{fig2}, the velocity profile of a relativistic rarefaction wave is no
longer linear as in the Newtonian case because of the relativistic velocity
addition formula. Comparing our results with the simulations of Hawley \etal
(1984b) and Laguna \etal (1993) for the same $\gamma=1.4$ shock tube, we note
that numerical inaccuracies due to the non-conservative formulation of
their energy equation clearly show up in their results for the relativistic
rest-mass density and specific internal energy. To investigate the
convergence properties of our numerical method, we performed a calculation of
the $\gamma=1.4$ shock tube with $10000$ particles and initial smoothing
length $h=0.1$ ($\approx20$ interactions per particle). As can be seen in
Figure \ref{fig3}, the numerical calculation covers the analytic solution
almost exactly, and the largest relative error is reduced to $\Delta
v=0.2\%$.

When the relativistic $\gamma$-factor of the shock tube is increased by
increasing the initial specific internal energy ratio, the region between the
leading shock front and the trailing contact discontinuity becomes extremely
thin and dense. Thus, without specially designed adaptive methods, it will be
impossible to resolve these Lorentz contracted shells of matter, which are
typical for relativistic fluid flows. In addition, the specific internal
energy at the contact discontinuity may become negative in the SPH
simulations. In order to avoid these difficulties, we consider now a
simplified problem of a single shock front without the presence of a contact
discontinuity and a nonlinear rarefaction wave.

\subsubsection*{b) Wall Shock Tests}

\begin{figure}[t]
\begin{center}
\begin{minipage}[t]{\figuresize}
\epsfxsize=\boxsize
\epsfbox{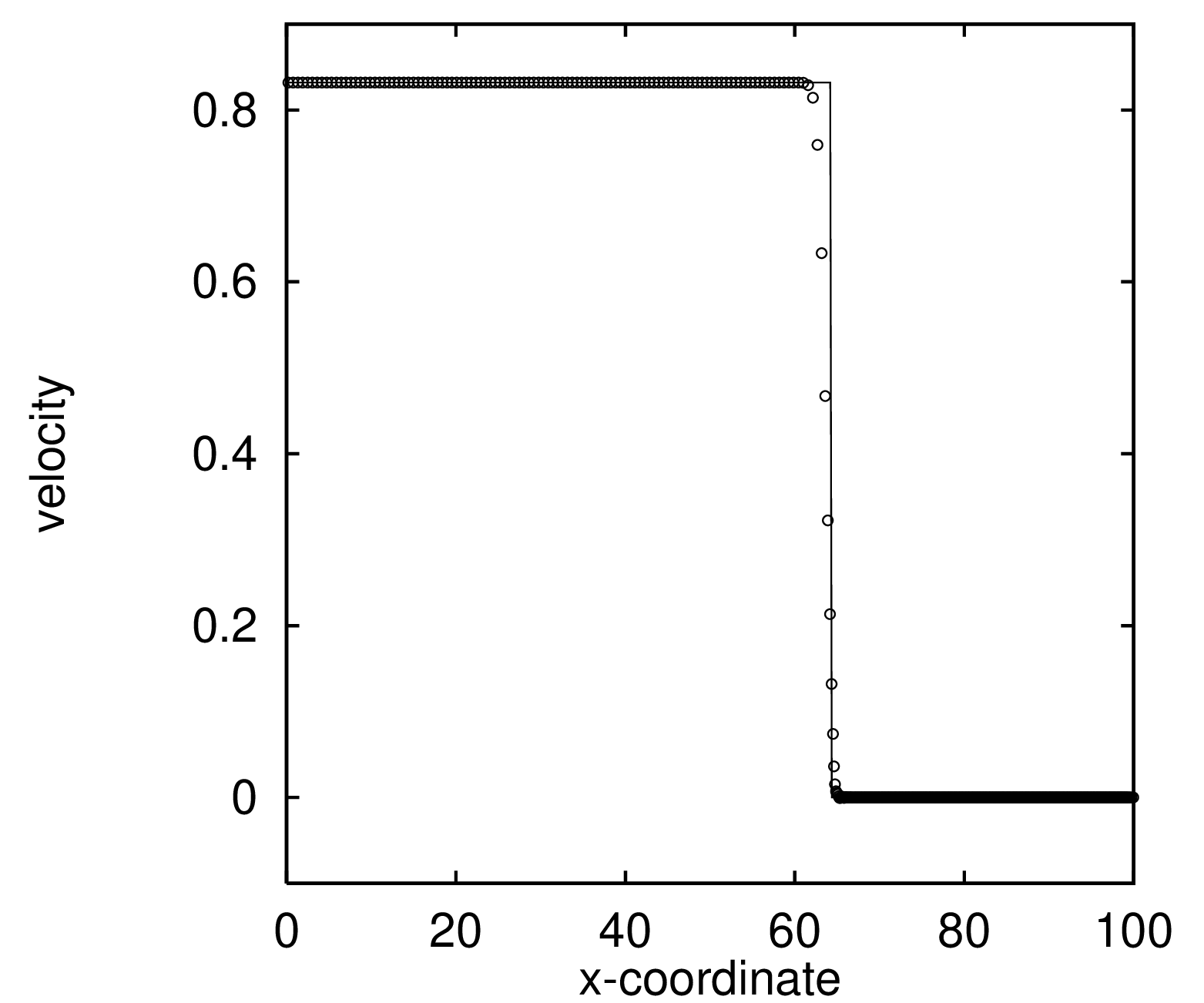}
\end{minipage}
\hspace{15mm}
\begin{minipage}[t]{\figuresize}
\epsfxsize=\boxsize
\epsfbox{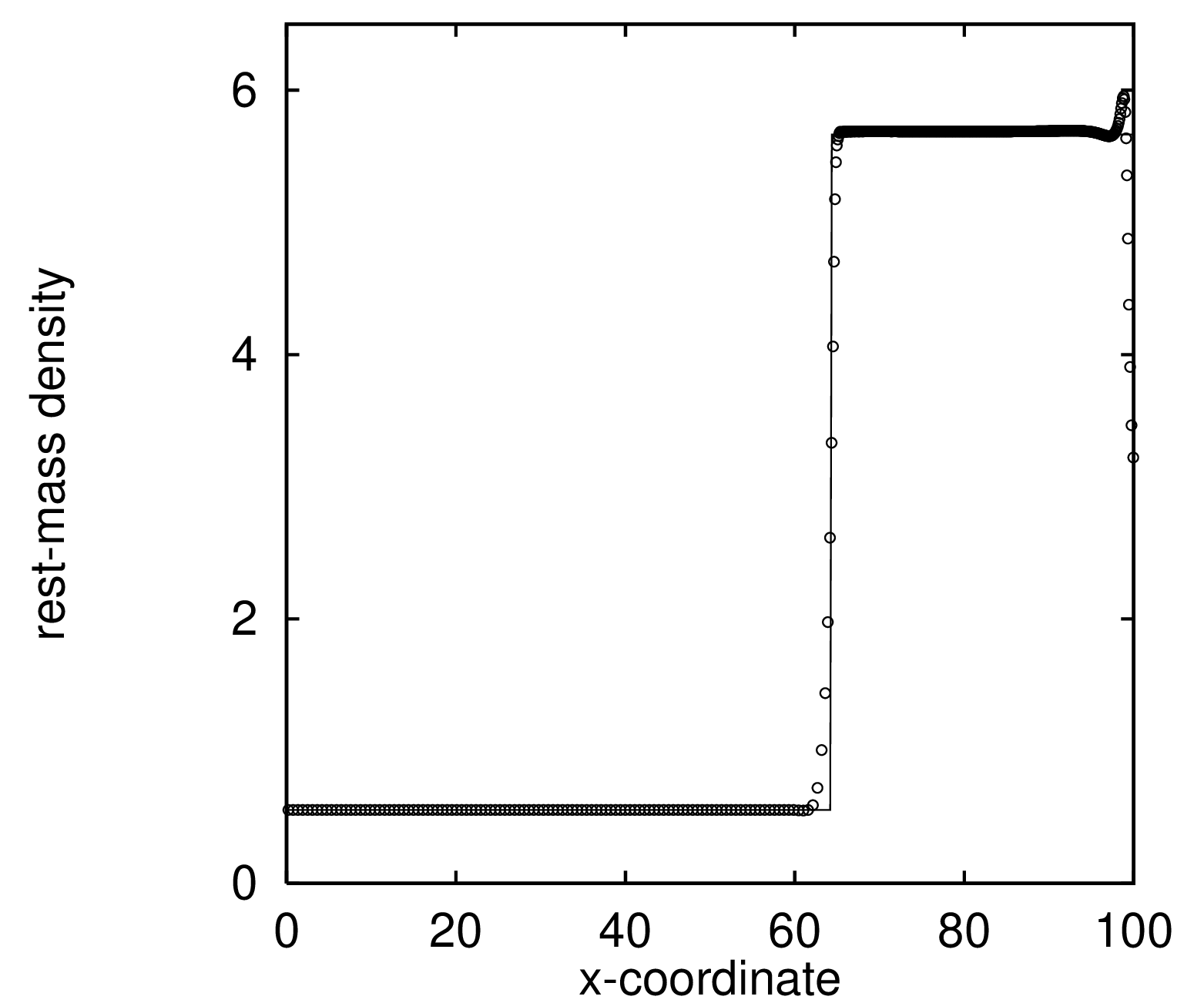}
\end{minipage}
\\
\vspace{10mm}
\begin{minipage}[t]{\figuresize}
\epsfxsize=\boxsize
\epsfbox{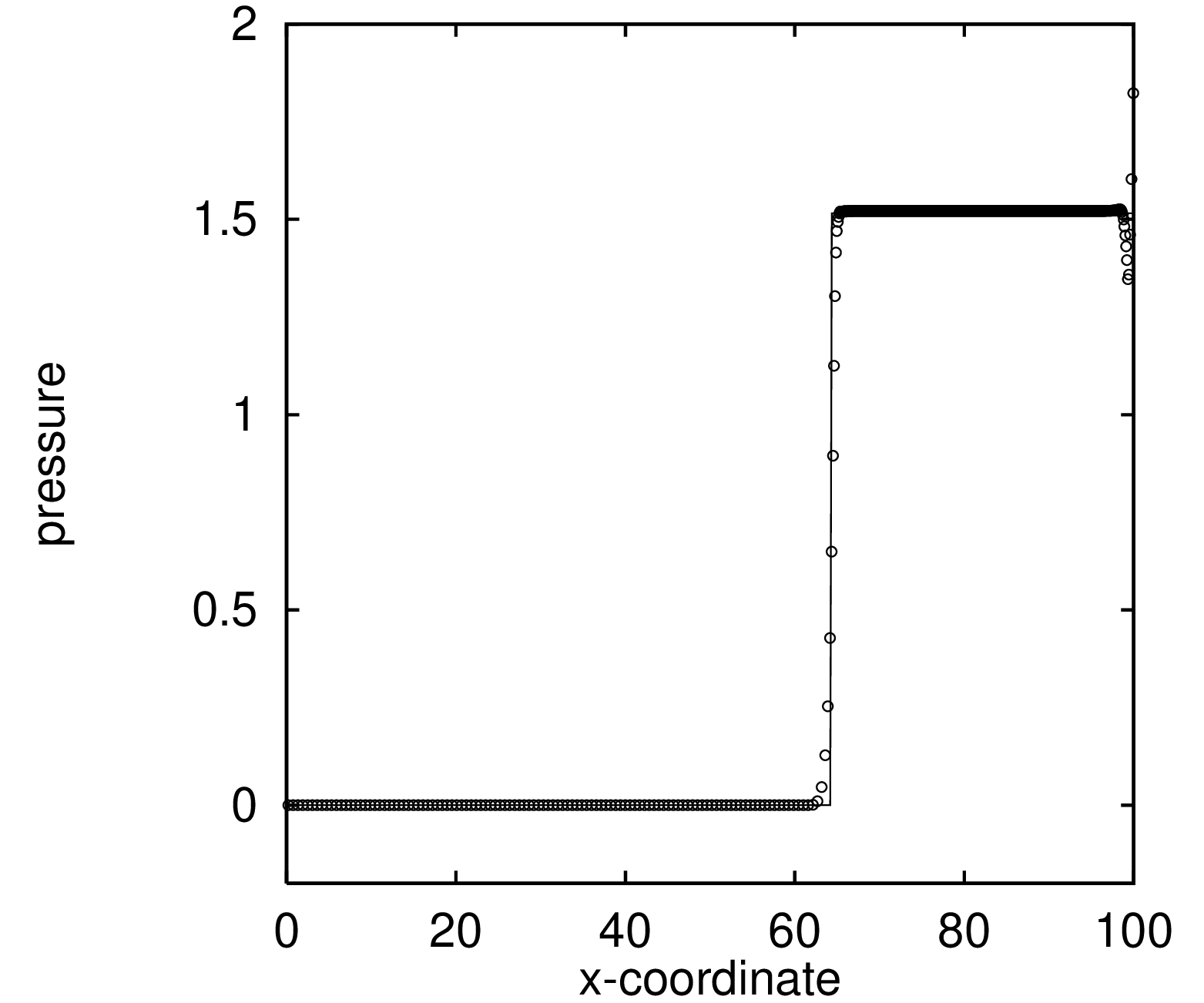}
\end{minipage}
\hspace{15mm}
\begin{minipage}[t]{\figuresize}
\epsfxsize=\boxsize
\epsfbox{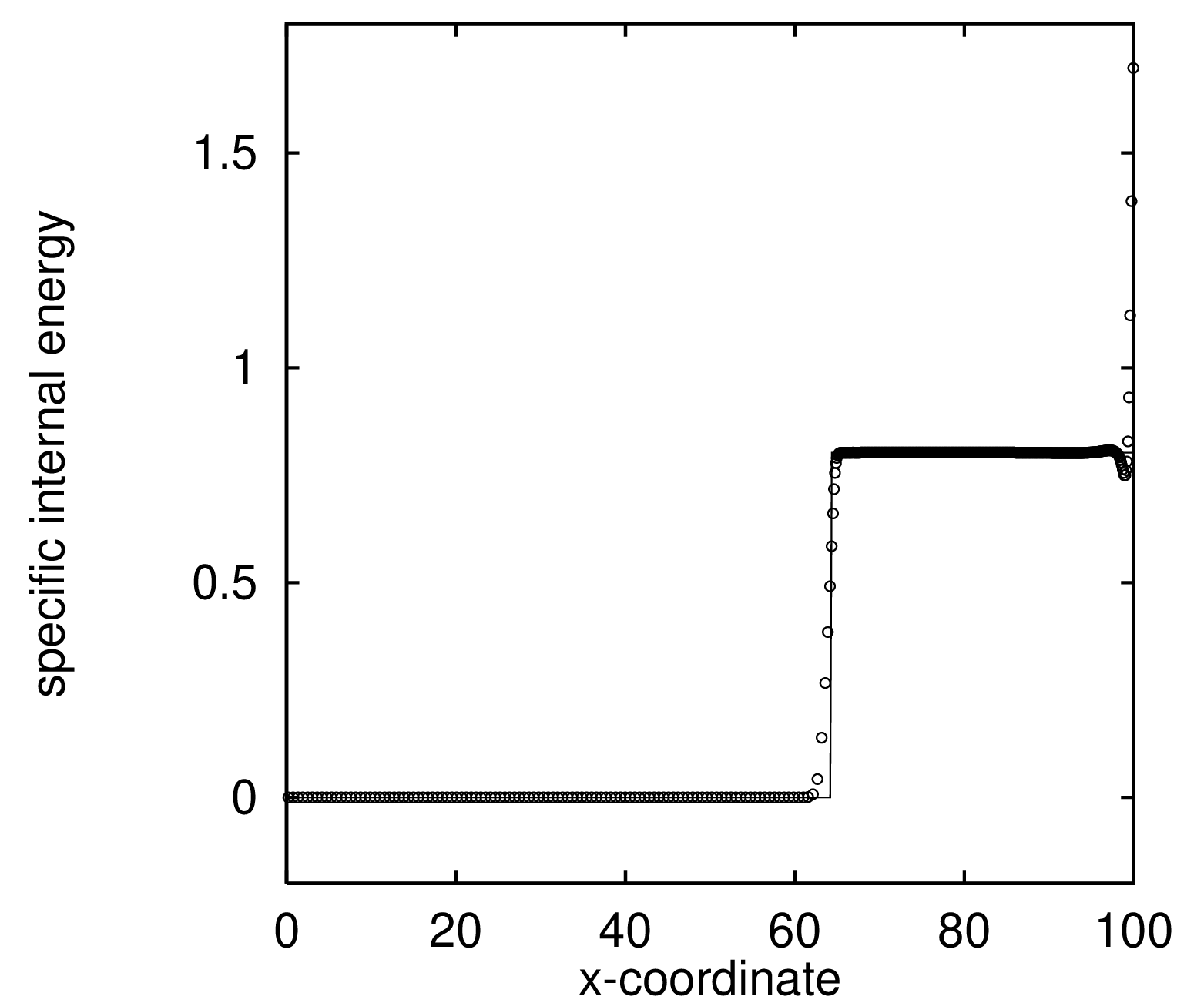}
\end{minipage}
\end{center}
\caption{\label{fig4} Numerical result of an SPH calculation with about $250$
particles (open circles) and analytic solution (solid line) for the
relativistic $\gamma_i=1.8$ wall shock problem. The post shock properties are
$\rho_p=5.66$, $p_p=1.52$, and $\varepsilon_p=0.803$, and the position and
velocity of the shock front are $x_s=64.3$ and $v_s=-0.178$ with
$\gamma_s=1.02$.}
\end{figure}

As a second test of our numerical method we modeled the wall shock problem of
a cold relativistically moving fluid flowing towards a solid wall. As the
fluid hits the wall, a shock front forms, which then travels upstream against
the incoming fluid producing a hot and dense post shock region of zero
velocity.

\begin{figure}[t]
\begin{center}
\begin{minipage}[t]{\figuresize}
\epsfxsize=\boxsize
\epsfbox{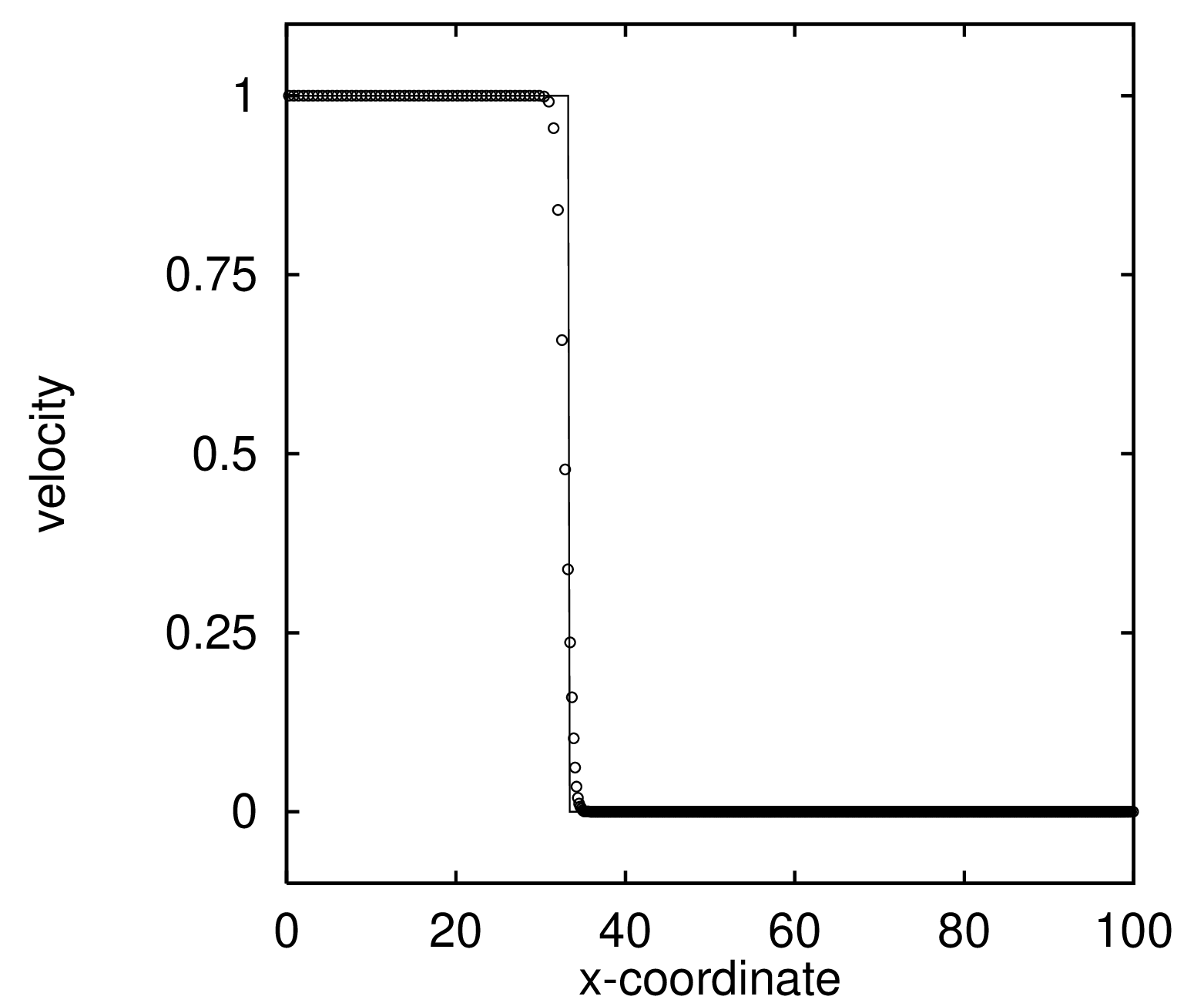}
\end{minipage}
\hspace{15mm}
\begin{minipage}[t]{\figuresize}
\epsfxsize=\boxsize
\epsfbox{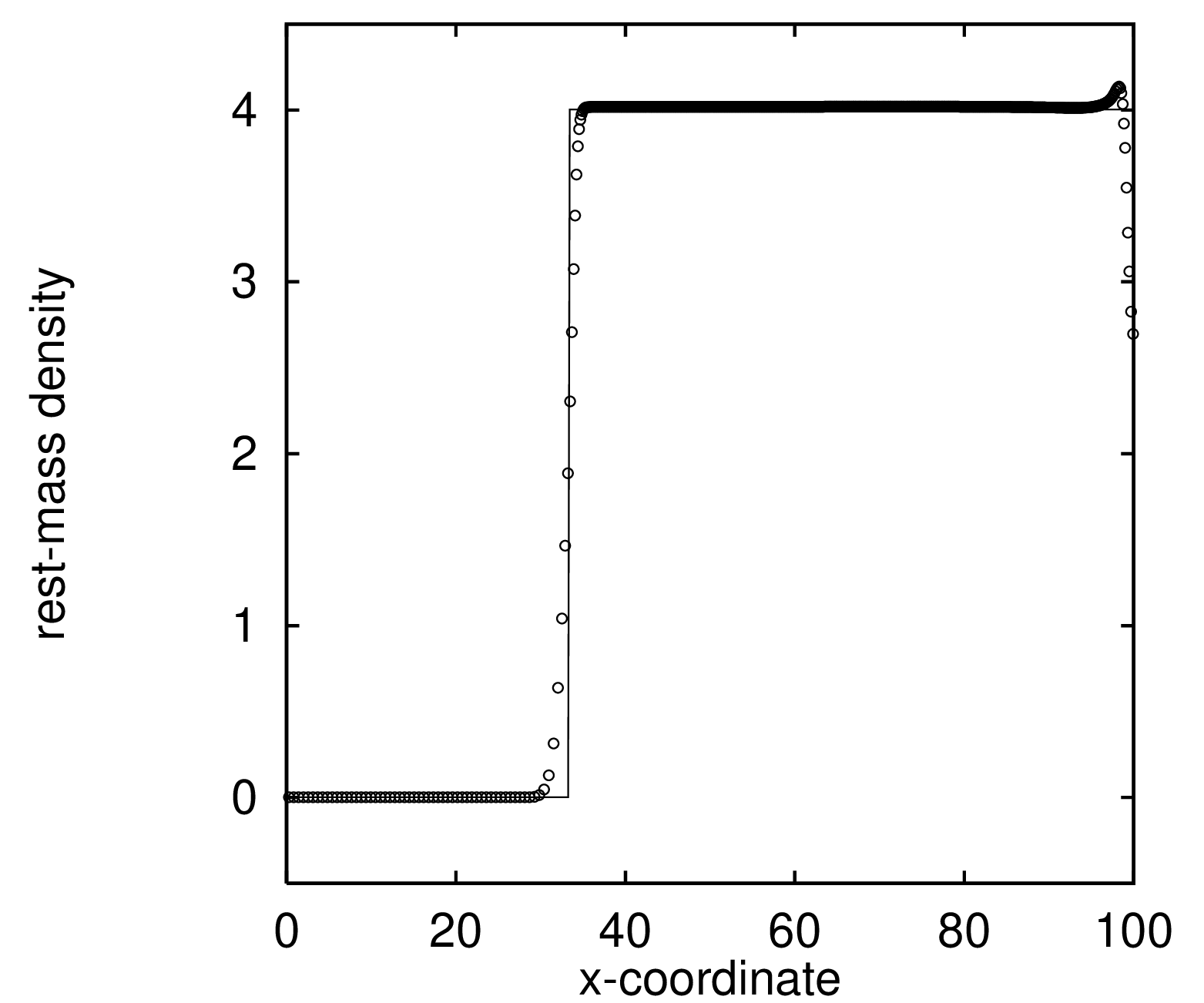}
\end{minipage}
\\
\vspace{10mm}
\begin{minipage}[t]{\figuresize}
\epsfxsize=\boxsize
\epsfbox{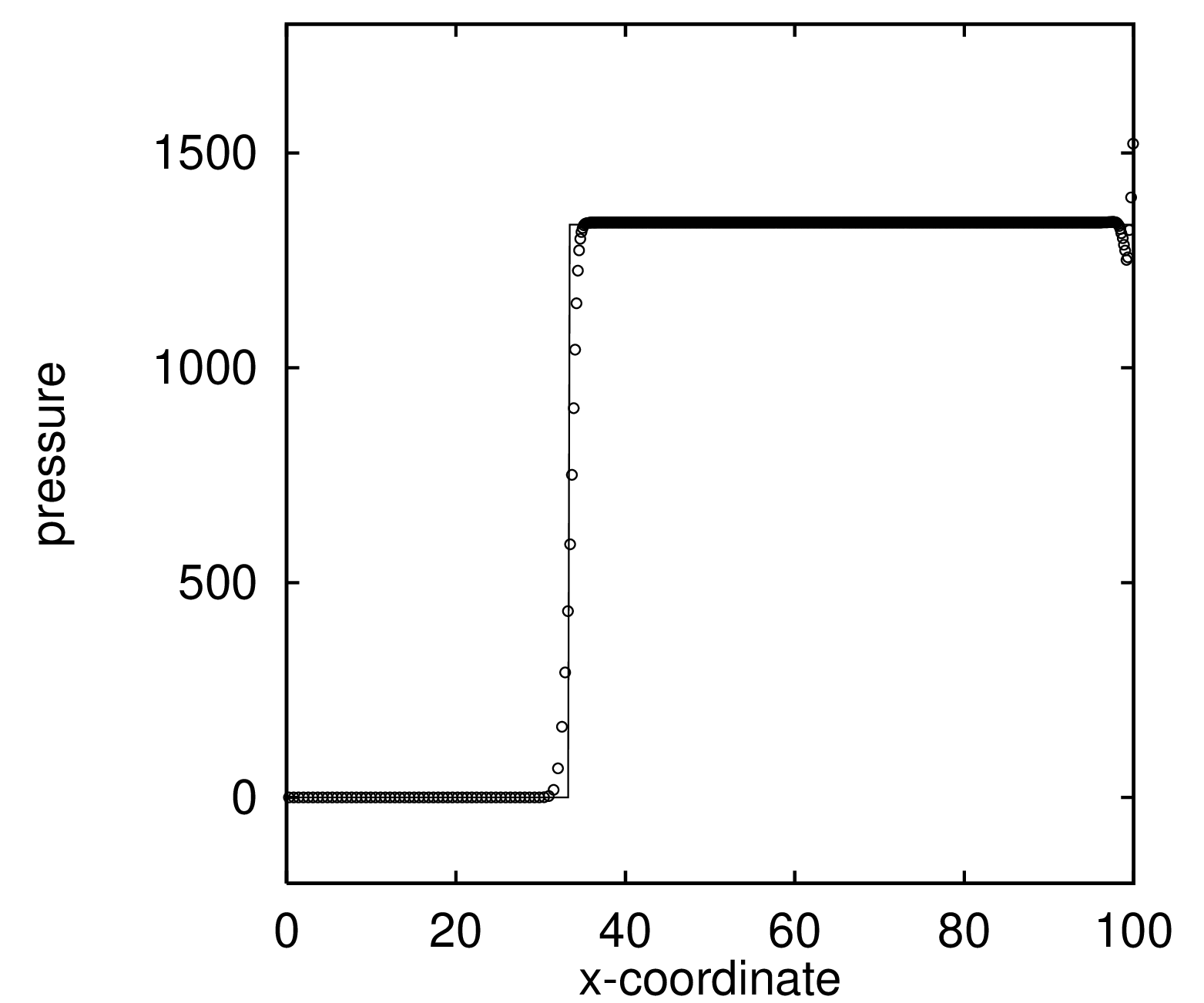}
\end{minipage}
\hspace{15mm}
\begin{minipage}[t]{\figuresize}
\epsfxsize=\boxsize
\epsfbox{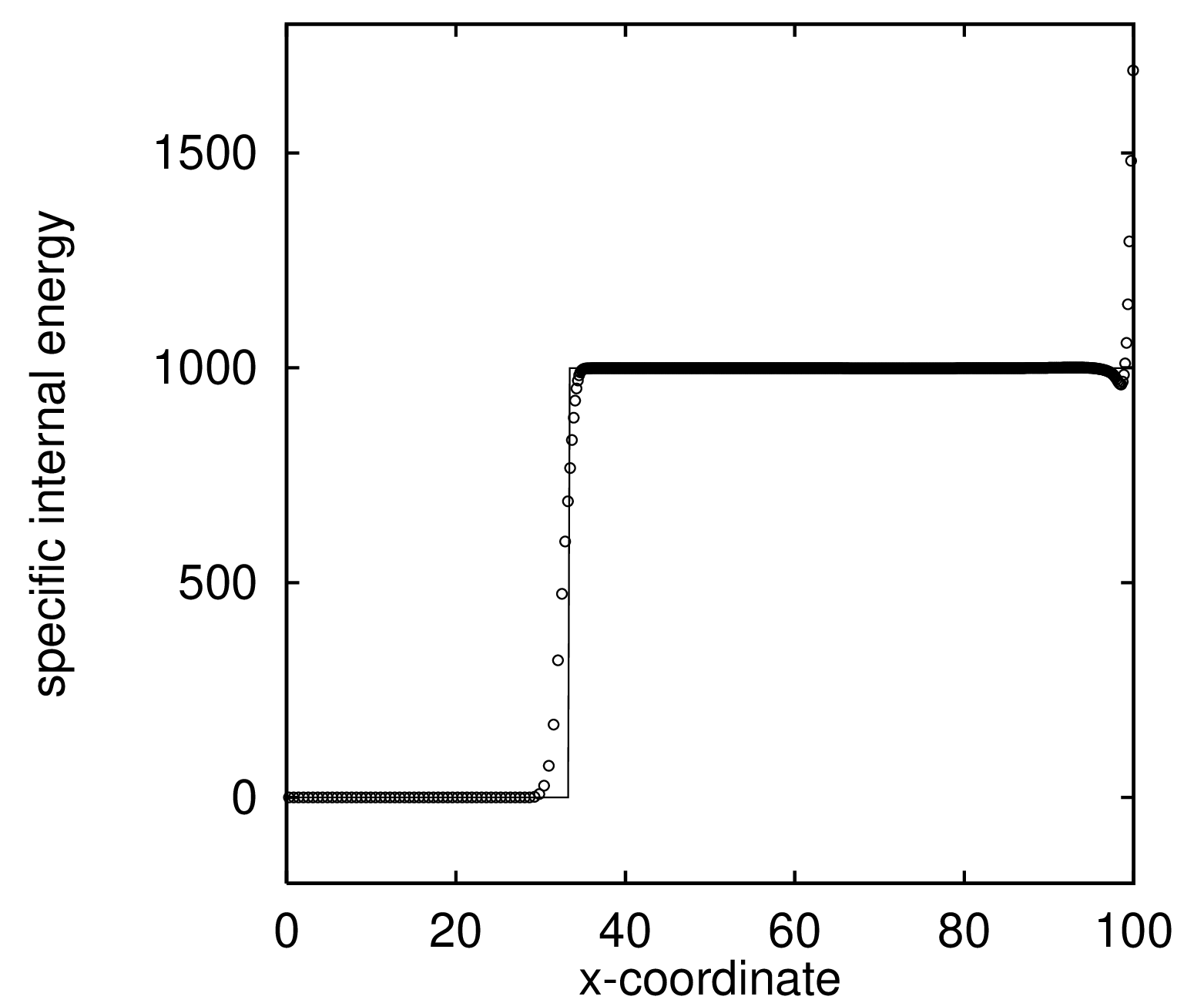}
\end{minipage}
\end{center}
\caption{\label{fig5} Numerical result of an SPH calculation with about
$250$ particles (open circles) and analytic solution (solid line) for the
ultra-relativistic $\gamma_i=1000$ wall shock problem. The post shock
properties are $\rho_p\approx4$, $p_p\approx4/3\times10^3$, and
$\varepsilon_p\approx1000$, and the position and velocity of the shock front
are $x_s\approx33.3$ and $v_s\approx-1/3$ with $\gamma_s\approx1.06$.}
\end{figure}

Figure \ref{fig4} shows the numerical result and the analytic solution of a
mildly relativistic wall shock for a gas with $\Gamma=4/3$ at time $t=200$
moving to the right with the reflecting wall at $x=100$. The uniform initial
fluid properties are $D_i=1$, $v_i=0.832$ with $\gamma_i=1.8$, and
$\varepsilon_i=10^{-5}$. Initially, particles of equal mass are uniformly
distributed in the simulation domain. At the location of the solid wall we
use reflecting boundary conditions. The simulation of Figure \ref{fig4}
showing about $250$ particles was performed with the initial smoothing length
$h=3$ ($\approx10$ interactions per particle) and artificial viscosity
parameters $\tilde\alpha=0.25$ and $\tilde\beta=0.5$. The rest-mass density
$\rho$, the thermodynamic pressure $p$, and the specific internal energy
$\varepsilon$ show a spike-like feature (see Fig.~\ref{fig4}), which is
known in the literature as ``wall heating'' (Norman \& Winkler 1986). The
largest relative error appears in the specific internal energy where
$\Delta\varepsilon=1.1\%$.

The results of an ultra-relativistic shock simulation are shown in Figure
\ref{fig5}. The initial velocity $v_i$ is increased to $v_i=0.9999995$ which
corresponds to a relativistic $\gamma$-factor of $\gamma_i=1000$. All other
parameters are identical with those of the previous wall shock calculation.
Neglecting the pre-shock specific internal energy $\varepsilon_i$, one can
show that in the ultra-relativistic limit $v_i\to1$ the post shock properties
$\rho_p$, $p_p$, $\varepsilon_p$, and the shock velocity $v_s$ are given by
$\rho_p=D_i\Gamma/(\Gamma -1)$, $p_p = \Gamma D_i\gamma_i$,
$\varepsilon_p=\gamma_i$, and $v_s=-(\Gamma-1)$. For the wall shock problem
of Figure \ref{fig5} we thus obtain $\rho_p\approx4$,
$p_p\approx4/3\times10^3$, $\varepsilon_p\approx1000$, and $v_s\approx-1/3$.
At the time $t=200$ the shock front has moved the distance
$|v_s|t\approx66.7$ to the left reaching the spatial position
$x_s\approx33.3$. The largest relative error in this case is $\Delta
\rho=1.1\%$.

\begin{figure}[t]
\begin{center}
\begin{minipage}[t]{\figuresize}
\epsfxsize=\boxsize
\epsfbox{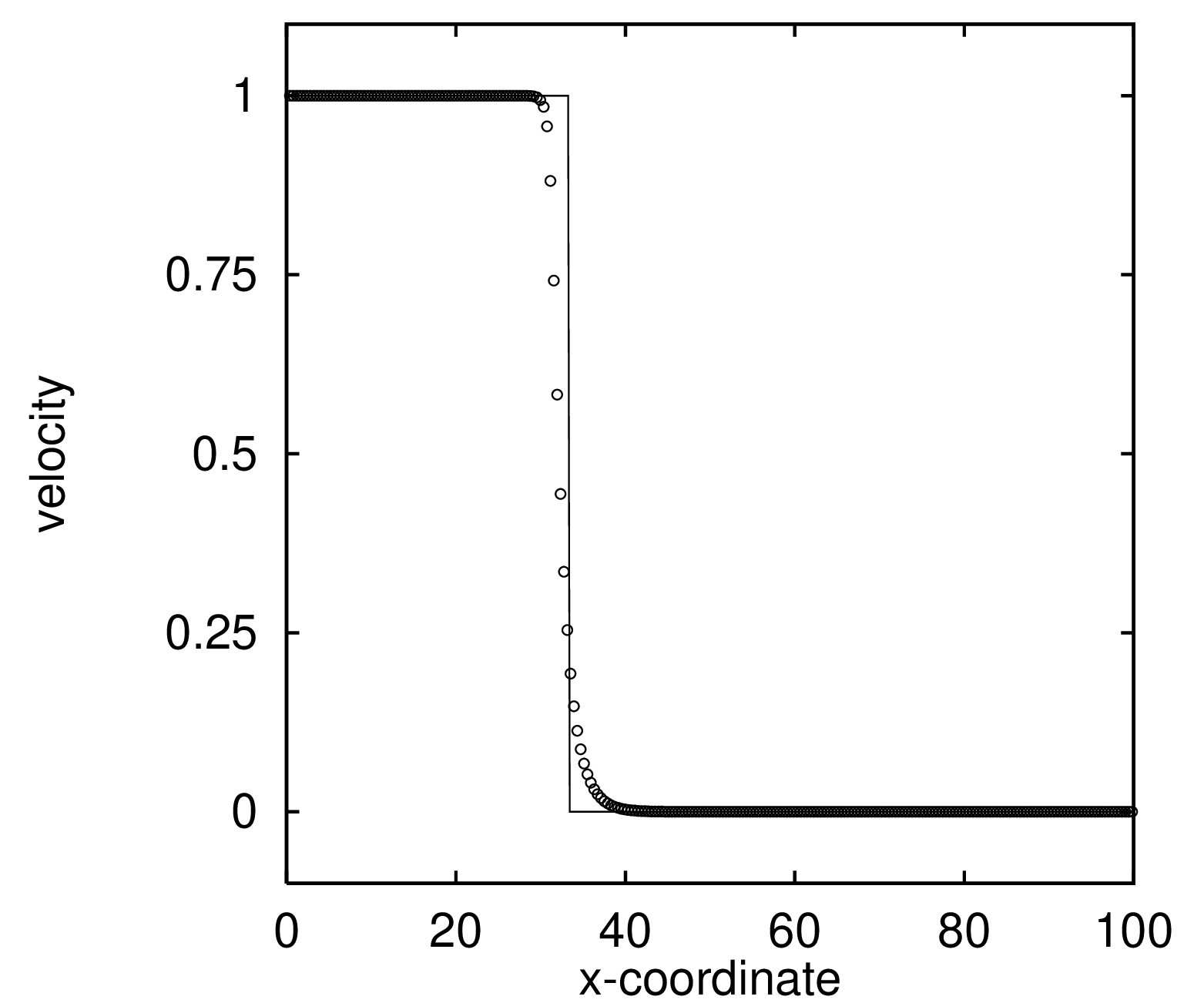}
\end{minipage}
\hspace{15mm}
\begin{minipage}[t]{\figuresize}
\epsfxsize=\boxsize
\epsfbox{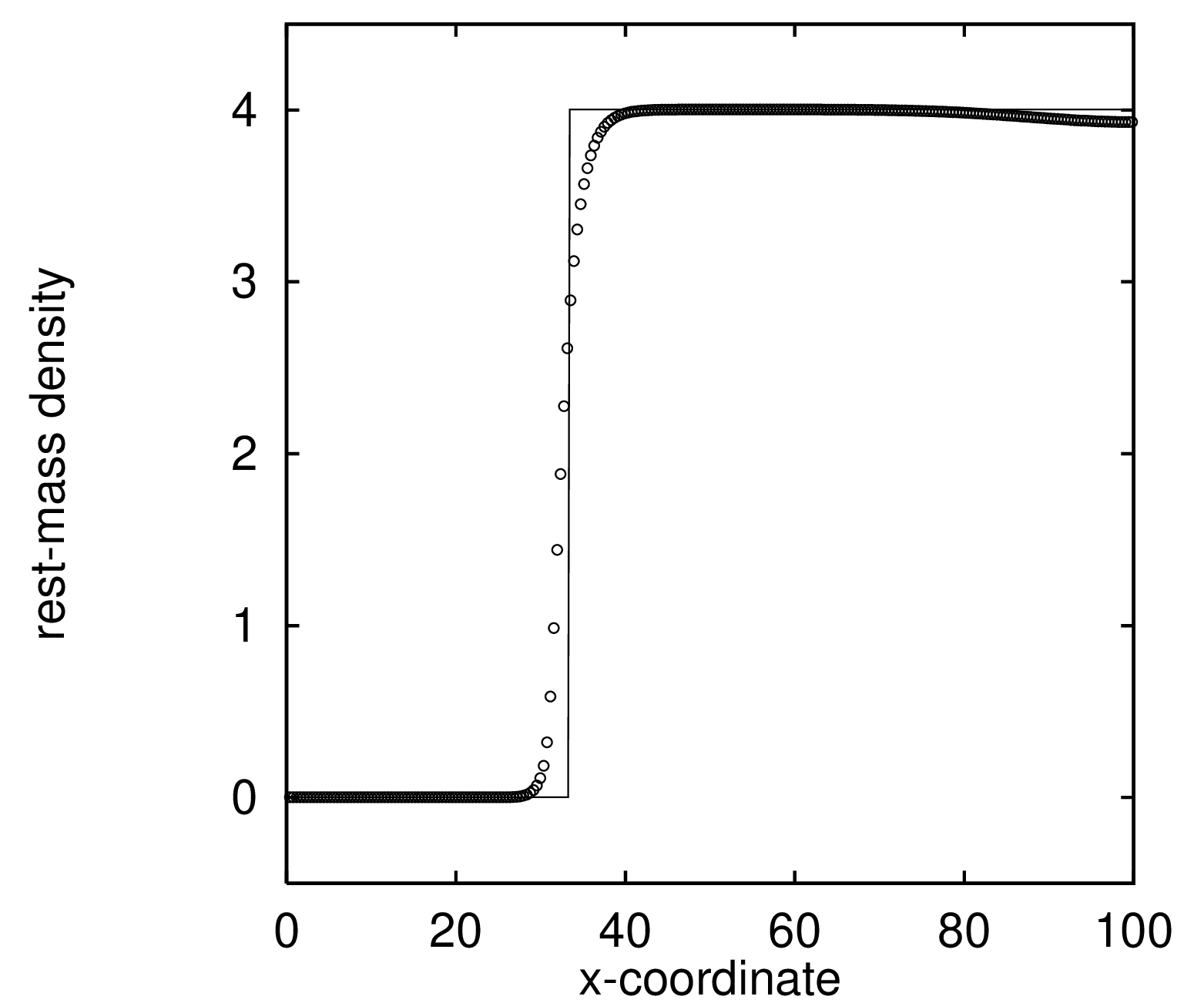}
\end{minipage}
\\
\vspace{10mm}
\begin{minipage}[t]{\figuresize}
\epsfxsize=\boxsize
\epsfbox{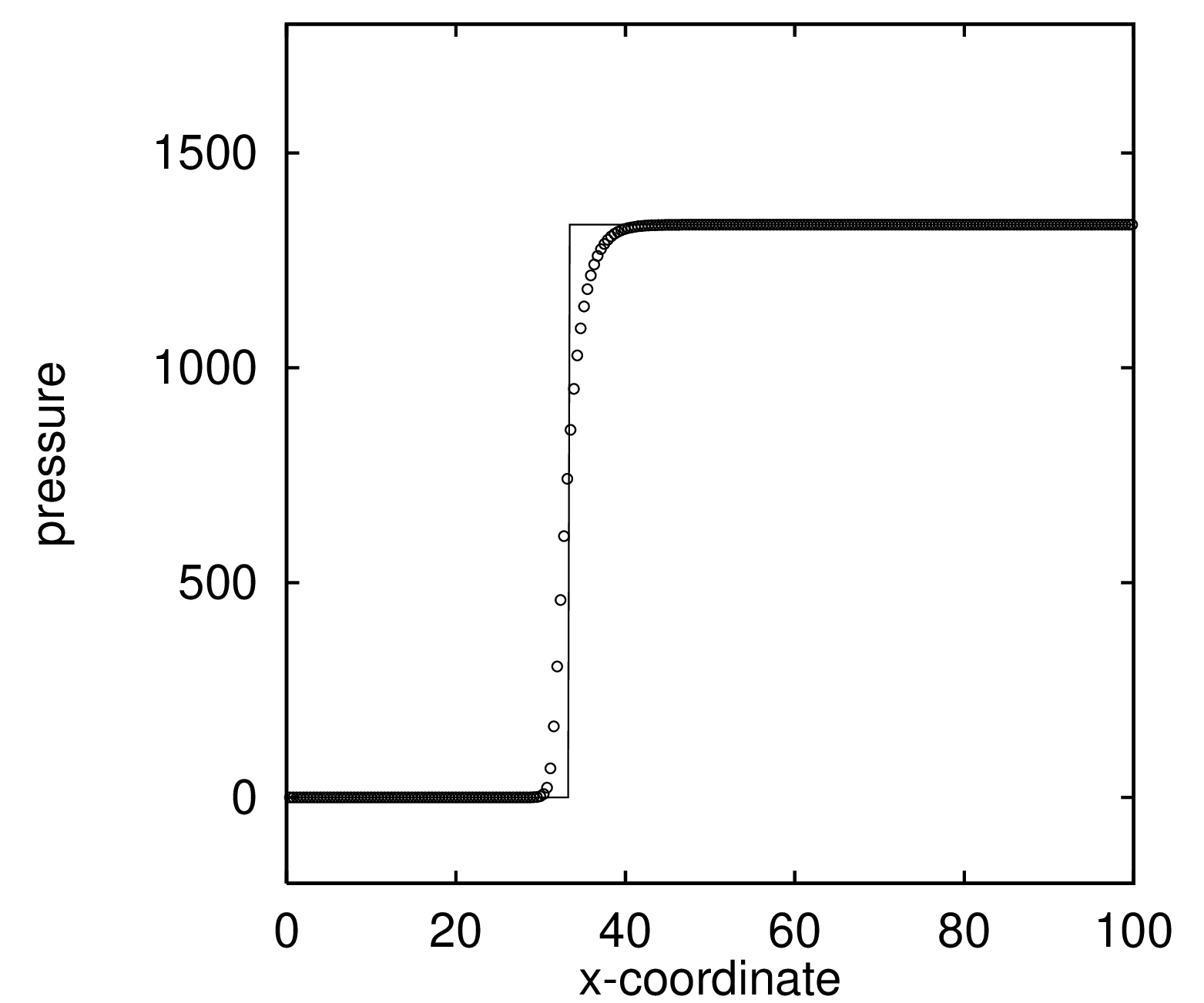}
\end{minipage}
\hspace{15mm}
\begin{minipage}[t]{\figuresize}
\epsfxsize=\boxsize
\epsfbox{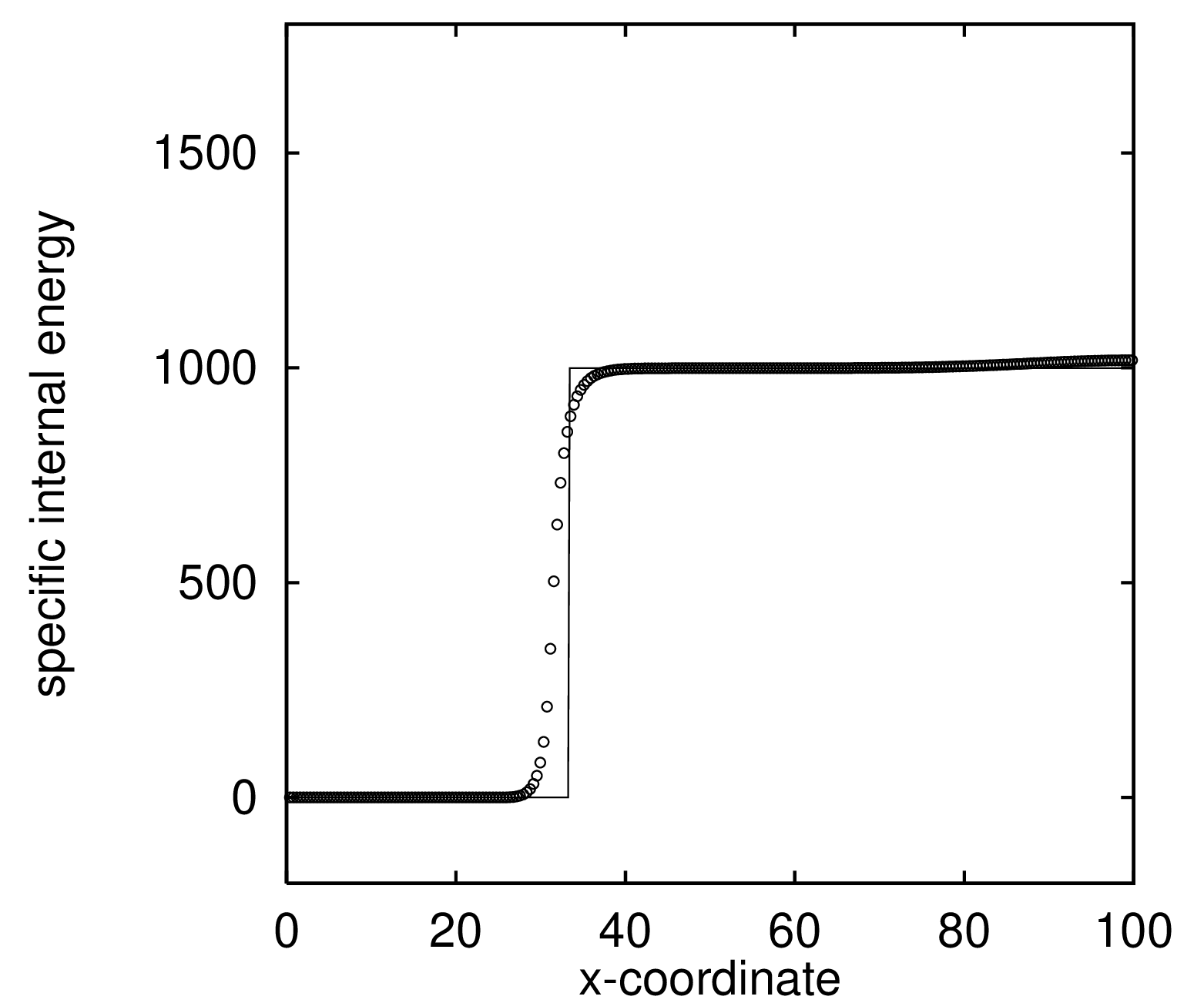}
\end{minipage}
\end{center}
\caption{\label{fig6} Numerical result of a finite difference calculation
with a grid of $250$ zones (open circles) and analytic solution (solid line)
for the ultra-relativistic $\gamma_i=1000$ wall shock problem. The post shock
properties are $\rho_p\approx4$, $p_p\approx4/3\times10^3$, and
$\varepsilon_p\approx1000$, and the position and velocity of the shock front
are $x_s\approx33.3$ and $v_s\approx-1/3$ with $\gamma_s\approx1.06$.}
\end{figure}

A calculation of a mildly relativistic wall shock problem (with
$\gamma_i=2.24$) using artificial viscosity was also performed by Hawley
\etal (1984b) with their time-explicit Eulerian finite difference code. They
tried several formulations of their non-conservative energy equation by
omitting and including various artificial viscous pressure terms. They found
that the $q\partial_t\gamma$ term in their energy equation is unstable even
for mildly relativistic wall shocks. This is verified by the calculations of
Norman \& Winkler (1986) using an implicit adaptive-mesh finite difference
method, which shows the requirement of an implicit technique to handle the
$q\partial_t\gamma$ term. The fact that our method is able to treat even
ultra-relativistic shocks without problems is entirely based on our
conservation formulation of the relativistic hydrodynamic equations where no
additional time derivatives of hydrodynamic variables appear such as the
$q\partial_t\gamma$ term. In order to verify that this numerical stability is
not restricted to the SPH method, we also performed numerical simulations of
the wall shock problem with a simple explicit finite difference upwind
scheme. Figure \ref{fig6} shows the numerical result and the analytic
solution for the $\gamma_i=1000$ wall shock at time $t=200$ using a grid of
$250$ zones, and the maximum relative error occurs for the specific internal
energy where $\Delta\varepsilon=2.2\%$.
\section{\label{summary}SUMMARY}
We have derived a fully Lagrangian conservative form of the general
relativistic equations of hydrodynamics for a perfect fluid with artificial
viscosity in a given arbitrary background spacetime. This has been achieved
by choosing suitable Lagrangian time evolution variables. These variables are
connected to the generic fluid variables rest-mass density $\rho$, 3-velocity
$\bar v^i$, and thermodynamic pressure $p$ through a set of nonlinear
algebraic equations. For an ideal gas, we have shown that these equations can
be reduced to a single fourth-order equation with a unique solution which can
be explicitly calculated in terms of various roots. For more complex
equations of state the solution of the above algebraic equations has to be
performed numerically, and the question of uniqueness is more complicated.
Using our Lagrangian formulation, we have developed a three-dimensional
general relativistic SPH code based on the standard SPH approach of
non-relativistic fluid dynamics. The important point is that all metric
factors from covariant derivatives have been absorbed in the definition of
the Lagrangian variables and thus no longer appear in the fluid equations. As
a result, the SPH kernels remain spherically symmetric and are the same for
all particles. This is an essential difference to the covariant SPH approach
of Kheyfets \etal (1990) using kernels defined in the local comoving frame of
the fluid, and to the relativistic SPH method of Laguna \etal (1993) where,
in curved spacetime, the kernels are anisotropic and have no translational
symmetry which leads to additional terms in the SPH equations. The
relativistic continuity equation (\ref{lagconti}), for example, contains no
source term, hence we can identify the relativistic rest-mass density $D^* =
\sqrt{\eta} \gamma \rho$ as the appropriate SPH density for smoothing the
equations.

In this paper we have restricted ourselves to the numerical simulation of two
different one-dimensional examples, i.e., the special relativistic shock tube
and the wall shock. An empirical error estimate is obtained from a comparison
of the numerical results with the corresponding analytic solutions. The SPH
calculations with $1000$ particles show a typical maximum relative error of
about $1\%$, and an increase of the particle number with a decreasing
smoothing length reduces this error in a uniform way. The wall-shock problem
can be solved without any numerical difficulties for very large
$\gamma$-factors (at least $\gamma = 1000$). The shock-tube case suffers from
a resolution problem if $\gamma$ is large because a thin shell of matter
builds up between the shock front and the contact discontinuity. This can
only be resolved with the help of additional adaptive methods which increase
the number of particles in this zone.

An important ingredient in our SPH formulation is the treatment of shock
structures by an artificial viscosity rather than using a Riemann solver.
This is very easy to implement even for the two- or three-dimensional case.
Introducing an artificial viscous pressure $q$ is considered as a purely
numerical tool that acts as a filter to smear out steep gradients in the
hydrodynamic functions and suppresses unphysical oscillations. Thus, there is
no need to use a covariant expression for $q$ in terms of velocity
derivatives as long as the time evolution and the jump conditions of shocks
are represented correctly. Since the jump conditions follow from the
conservation of mass, momentum, and energy across the shock, it is obvious
that a conservative form of the relativistic hydrodynamic equations is
important for handling discontinuities in relativistic fluid dynamics
numerically. The simulation of relativistic flows with large $\gamma$-factors
including shocks is the traditional domain of HRSC methods. However, with our
formulation of the general relativistic hydrodynamic equations it is possible
to model such flows using an artificial viscosity independent of the
underlying numerical method, i.e., SPH or finite difference schemes.
\acknowledgments
\section*{ACKNOWLEDGEMENTS}
We want to thank P. Laguna for helpful discussions and for encouraging us to
publish our results. This work is part of the project A7 of the
Sonderforschungsbereich SFB~382 funded by the Deutsche Forschungsgemeinschaft
(DFG).
\newpage
\appendix
\section{APPENDIX}
In section \ref{hydroeq} we have derived equation (\ref{gamzeroeq}) from
which the relativistic $\gamma$-factor can be calculated analytically as the
root of a polynomial of degree four for given artificial viscous pressures
$q$. Equation (\ref{gamzeroeq}) is restricted to an ideal-gas equation of
state (\ref{idealstateeq}). We will now show the existence and uniqueness of
the solution of equation (\ref{gamzeroeq}) for all allowed values of $G$,
$S^2$, and $\tilde E$.

First, we rewrite equation (\ref{gamzeroeq}) by substituting $\gamma =
1+\delta$, $\delta\in[0,\infty)$ and obtain
\begin{eqnarray}
\label{deltazeroeq}
0 & = & \underbrace{\left(S^2 - \tilde E^2\right)}_{a_4}\delta^4 + 
\underbrace{2\left(2S^2 - 2\tilde E^2 + G\tilde E\right)}_{a_3}\delta^3
\nonumber\\
&&{}+\underbrace{\left(6S^2 - 5\tilde E^2 + 6G\tilde E - 2GS^2 -
G^2\right)}_{a_2}\delta^2\nonumber\\
&&{}+\underbrace{2\left(2S^2 - \tilde E^2 + 2G\tilde E - 2GS^2 -
G^2\right)}_{a_1}\delta + \underbrace{S^2(1-G)^2}_{a_0} \enspace .
\end{eqnarray}
With the degree of freedom $f\geq3$ the ideal-gas adiabatic constant
$\Gamma=1+2/f$ and the variable $G=1-1/\Gamma$ lie in the range 
\begin{equation}
\label{gammagrange}
1 <\Gamma\leq \frac{5}{3} ~~,~~~~
0 <G\leq \frac{2}{5} \enspace .
\end{equation}
From equations (\ref{hqrho}) and (\ref{ssquare}) the variables $\tilde E$
and $S^2$ are given by
\begin{equation}
\label{etildessquare}
\tilde E = \left(\gamma - \frac{G}{\gamma}\right)\tilde w + \frac{G}{\gamma}
~~,~~~~ S^2 = \tilde w^2\left(\gamma^2 - 1\right)\geq0 \enspace ,
\end{equation}
where we have defined $\tilde w = w+q/\rho\geq1$. Using the relations
(\ref{gammagrange}) and the expressions (\ref{etildessquare}), we obtain for
the coefficient $a_4$ in equation (\ref{deltazeroeq})
\begin{eqnarray}
\label{rela4}
a_4 & = & S^2 - \tilde E^2\nonumber\\
& = & -\left(1-2G+\frac{G^2}{\gamma^2}\right)\tilde w^2 -
2G\left(1-\frac{G}{\gamma^2}\right)\tilde w - \frac{G^2}{\gamma^2} < 0
\enspace . 
\end{eqnarray}
Thus, $S^2$ is limited to the range
\begin{displaymath}
0\leq S^2<\tilde E^2 \enspace .
\end{displaymath}

\subsubsection*{a) Existence}

With the coefficients $a_4<0$ and $a_0=S^2/\Gamma^2\geq 0$ equation
(\ref{deltazeroeq}) has at minimum one positive root of $\delta$.

\subsubsection*{b) Uniqueness}

To show the uniqueness of the solutions of equation (\ref{deltazeroeq}) for
$\delta$, we investigate  the changes of signs of the coefficients $a_3$,
$a_2$, and $a_1$ depending on the variable $S^2$:
\begin{eqnarray*}
a_3<0 & \Leftrightarrow & S^2<\tilde E^2 - \frac{1}{2}G\tilde
E =: S_3^2 \enspace ,\nonumber\\
a_2<0 & \Leftrightarrow & S^2<\frac{1}{2(3 - G)}\left(5\tilde E^2 - 6G\tilde E
+ G^2\right) =: S_2^2 \enspace , \quad \mbox{and}\nonumber\\
a_1<0 & \Leftrightarrow & S^2<\frac{1}{2(1 - G)}\left(\tilde E -
G\right)^2 =: S_1^2 \enspace .
\end{eqnarray*}
Using
\begin{displaymath}
\tilde E - G = \left(\tilde w - 1\right)\gamma\left(1 -
\frac{G}{\gamma^2}\right) + \gamma - G > 0
\end{displaymath}
and the relations (\ref{gammagrange}), one can show that
\begin{eqnarray}
\label{rels1}
S_1^2 & > & 0 \enspace ,\\ 
\label{rels1s2}
S_2^2 - S_1^2 & = & \frac{\Gamma}{1 + 2\Gamma}\left[(2 - \Gamma)\tilde E^2 +
2(\Gamma - 1)G\tilde E - \Gamma G^2\right]\nonumber\\
& > & \frac{\Gamma G^2}{1 + 2\Gamma}\left[(2 - \Gamma) + 2(\Gamma - 1) -
\Gamma\right] = 0 \enspace , \quad \mbox{and}\\
\label{rels2s3}
S_2^2 & = & \frac{1}{2(3 - G)}\left(5\tilde E^2 - 6G\tilde E +
G^2\right)\nonumber\\
& < & \frac{1}{2(3 - G)}\left(5\tilde E^2 - 6G\tilde E + G\tilde
E\right)\nonumber\\
& = & \frac{5}{2(3 - G)}\left(\tilde E^2 - G\tilde E\right)<\tilde E^2 -
G\tilde E < S_3^2 \enspace .
\end{eqnarray}
The relations (\ref{rels1}), (\ref{rels1s2}), and (\ref{rels2s3}) lead to
\begin{equation}
\label{rels1s2s3}
0 < S_1^2 < S_2^2 < S_3^2 \enspace .
\end{equation}

i) For $a_0=S^2/\Gamma^2>0$ or $S^2>0$, respectively, we obtain from the
relations (\ref{rela4}) and (\ref{rels1s2s3}) the following table of signs
for the coefficients $a_4$, $a_3$, $a_2$, $a_1$, and $a_0$ depending on
$S^2$:
\begin{center}
\begin{tabular}{|c|ccccc|}
\hline
$S^2$ & $a_4$ & $a_3$ & $a_2$ & $a_1$ & $a_0$\\
\hline
\vspace{-5mm}&  &  &  &  &  \\
$S_3^2<S^2<{\tilde E}^2$ & $-$ & $+$ & $+$ & $+$ & $+$ \\
$S^2=S_3^2$ & $-$ & $0$ & $+$ & $+$ & $+$ \\
$S_2^2<S^2<S_3^2$ & $-$ & $-$ & $+$ & $+$ & $+$ \\
$S^2=S_2^2$ & $-$ & $-$ & $0$ & $+$ & $+$ \\
$S_1^2<S^2<S_2^2$ & $-$ & $-$ & $-$ & $+$ & $+$ \\
$S^2=S_1^2$ & $-$ & $-$ & $-$ & $0$ & $+$ \\
$0<S^2<S_1^2$ & $-$ & $-$ & $-$ & $-$ & $+$ \\
\hline
\end{tabular}
\end{center}
For all values of $S^2\in(0,\tilde E^2)$ the series of coefficients
$a_4$, $a_3$, $a_2$, $a_1$, and $a_0$ has exactly one change of sign.
Therefore, in $\delta\in(0,\infty)$ there exists only one root of
equation (\ref{deltazeroeq}) for $\delta$.

\placetable{tab1}

ii) For $a_0=S^2/\Gamma^2=0$ or $S^2=0$, respectively, equation
(\ref{deltazeroeq}) has the root $\delta=0$. With the relations (\ref{rela4})
and (\ref{rels1s2s3}) the coefficients $a_4$, $a_3$, $a_2$, and $a_1$ are all
less than zero. Thus, $\delta=0$ is the only positive root of equation
(\ref{deltazeroeq}) for $\delta$.

To summarize, with the restriction of the ideal-gas equation of state
(\ref{idealstateeq}) a solution of equation (\ref{deltazeroeq}) for $\delta$
or of equation (\ref{gamzeroeq}) for $\gamma$, respectively, exists and is
unique. Thus, the fluid variables rest-mass density $\rho$, 3-velocity $\bar
v^i$, and thermodynamic pressure $p$ can be calculated analytically in a
unique way from the variables $D^*$, $S_i$, $E$, and $q$ from equations
(\ref{gamzeroeq}), (\ref{densityvar}), (\ref{idealstateeq}), (\ref{hqrho}),
and (\ref{momentumvar}).

\newpage
\begin{references}

\reference {} Arnowitt, R., Deser, S., \& Misner, C.~W.~1962, in Gravitation,
ed.~L.~Witten (New York: Wiley), 227

\reference{} Banyuls, F., Font, J.~A., Ib\'a\~nez,
J.~$\mbox{M}^{\mbox{\scriptsize\b a}}$., Marti,
J.~$\mbox{M}^{\mbox{\scriptsize\b a}}$., \& Miralles, J.~A.~1997, \apj, 476,
221

\reference{} Falle, S.~A.~E.~G., \& Komissarov, S.~S. 1996, \mnras, 278, 586

\reference{} Font, J.~A., Ib\'a\~nez, J.~$\mbox{M}^{\mbox{\scriptsize\b
a}}$., Marquina, A., \& Marti, J.~$\mbox{M}^{\mbox{\scriptsize\b
a}}$.~1994, \aap, 282, 304

\reference{} Gingold, R.~A., \& Monaghan, J.~J.~1977, \mnras, 181, 375

\reference{} Hawley, J.~F., Smarr, L.~L., \& Wilson, J.~R.~1984a, \apj, 277,
296

\reference{} Hawley, J.~F., Smarr, L.~L., \& Wilson, J.~R.~1984b, \apjs, 55,
211

\reference{} Kheyfets, A., Miller, W.~A., \& Zurek, W.~H.~1990, \prd,
41, 451

\reference{} Komissarov, S.~S.~1999, \mnras, 303, 343

\reference{} Laguna, P., Miller, W.~A., \& Zurek, W.~H.~1993, \apj,
404, 678

\reference{} Landau, L.~D., \& Lifschitz, E.~M.~1991, Hydrodynamik (Berlin:
Akademie Verlag)

\reference{} LeVeque, R.~J.~1997, in Computational Methods for Astrophysical
Fluid Flow, lecture notes 1997/Saas-Fee Advanced Course 27, eds.~O.~Steiner
and A.~Gautschy (Berlin: Springer Verlag), 1

\reference{} Lucy, L.~B.~1977, \aj, 82, 1013

\reference{} Marti, J.~$\mbox{M}^{\mbox{\scriptsize\b a}}$., \& M\"uller,
E.~1994, J.~Fluid Mech., 258, 317

\reference{} McKee, C.~R., \& Colgate, S.~A.~1973, \apj, 181, 903

\reference{} Misner, C.~W., Thorne, K.~S., \& Wheeler, J.~A.~1973,
Gravitation (San Francisco: Freeman \& Co)

\reference{} Monaghan, J.~J.~1992, \araa, 30, 543

\reference{} Monaghan, J.~J., \& Gingold, R.~A.~1983, J.~Comput.~Phys., 52,
374

\reference{} Monaghan, J.~J., \& Lattanzio, J.~C.~1985, \aap, 149, 135

\reference{} von Neumann, J., \& Richtmyer, R.~D.~1950, J.~Appl.~Phys., 21,
232

\reference{} Norman, M.~L., \& Winkler, K.-H.~A.~1986, in Astrophysical
Radiation Hydrodynamics, eds.~M.~L.~Norman and K.-H.~A.~Winkler (Dordrecht:
Reidel), 449

\reference{} Pons, J.~A., Font, J.~A., Ib\'a\~nez,
J.~$\mbox{M}^{\mbox{\scriptsize\b a}}$., Marti,
J.~$\mbox{M}^{\mbox{\scriptsize\b a}}$., \& Miralles, J.~A.~1998, \aap, 339,
638

\reference{} Romero, J.~V., Ib\'a\~nez, J.~$\mbox{M}^{\mbox{\scriptsize\b
a}}$., Marti, J.~$\mbox{M}^{\mbox{\scriptsize\b a}}$., \& Miralles,
J.~A.~1996, \apj, 462, 839

\reference {} Taub, A.~H.~1948, Phys.~Rev., 74, 328

\reference{} Wen, L., Panaitescu, A., \& Laguna, P.~1997, \apj, 486, 919

\end{references}
\end{document}